\documentclass[twocolumn]{aastex631}

\usepackage{xspace}
\usepackage[nolist]{acronym}
\usepackage{graphicx}
\usepackage{subfigure}
\providecommand{\acrolowercase}[1]{\lowercase{#1}}

\newcommand{\pycbc}{PyCBC\xspace}
\newcommand{\pcl}{PyCBC Live\xspace}

\newcommand{\pastro}{$p_\mathrm{astro}$\xspace}

\begin{acronym}
\acro{TPR}[TPR]{True Positive Rate}
\acro{FPR}[FPR]{False Positive Rate}
\acro{BNS}[BNS]{binary neutron star}
\acro{NSBH}[NSBH]{neutron star-black hole}
\acro{MDC}[MDC]{Mock Data Challenge}
\acro{EoS}[EoS]{equation of state}
\acro{NS}[NS]{neutron star}
\acro{GW}[GW]{gravitational wave}
\acro{LIGO}[LIGO]{Laser Interferometer \acs{GW} Observatory}
\acro{aLIGO}[aLIGO]{Advanced \acs{LIGO}}
\acro{AdVirgo}[AdVirgo]{Advanced Virgo}
\acro{O4}[O4]{\ac{aLIGO}'s, \ac{AdVirgo}'s and KAGRA's fourth observing run}
\acro{LVK}[LVK]{LIGO-Virgo-KAGRA}
\acro{O1}[O1]{\ac{aLIGO}'s first observing run}
\acro{O2}[O2]{\ac{aLIGO}'s and \ac{AdVirgo}'s second observing run}
\acro{O3}[O3]{\ac{aLIGO}'s and \ac{AdVirgo}'s third observing run}
\acro{oLIB}[\acrolowercase{o}LIB]{Omicron+\acl{LIB}}
\acro{KAGRA}[KAGRA]{KAmioka GRAvitational\nobreakdashes-wave observatory}
\acro{GraceDB}[\texttt{GraceDB}]{GRAvitational-wave Candidate Event DataBase}
\acro{LLAI}[LLAI]{Low-Latency Alert Infrastructure}
\acro{CBC}[CBC]{compact binary coalescence}
\acro{BBH}[BBH]{binary black hole}
\acro{FAR}[FAR]{false alarm rate}
\acro{GstLAL}[GstLAL]{GStreamer LIGO Scientific Collaboration Algorithm Library}
\acro{SVD}[SVD]{singular value decomposition}
\acro{SNR}[SNR]{signal\nobreakdashes-to\nobreakdashes-noise ratio}
\acro{FGMC}[FGMC]{Farr, Gair, Mandel and Cutler formalism of Poisson Mixture Model}
\acro{MBTA}[MBTA]{Multi-Band Template Analysis}
\acro{SPIIR}[SPIIR]{Summed Parallel Infinite Impulse Response}
\acro{GPU}[GPU]{Graphical Processing Units}
\acro{cWB}[\acrolowercase{c}WB]{Coherent WaveBurst}
\acro{GCN}[GCN]{General Coordinates Network}
\acro{SCiMMA}[SCiMMA ]{Scalable CyberInfrastructure for Multi-Messenger Astrophysics}
\acro{BAYESTAR}[\texttt{BAYESTAR}]{BAYESian TriAngulation and Rapid localization}
\acro{FITS}[FITS]{Flexible Image Transport System}
\acro{HEALPix}[HEALP\acrolowercase{ix}]{Hierarchical Equal Area isoLatitude Pixelization}
\acro{MOC}[MOC]{Multi-Order Coverage}
\acro{EM}[EM]{electromagnetic}
\acro{NS}[NS]{neutron star}
\acro{BH}[BH]{black hole}
\acro{LLOID}[LLOID]{Low Latency Online Inspiral Detection}
\acro{IIR}[IIR]{infinite impulse response}
\acro{ToO}[ToO]{target of opportunity}
\acro{PE}[PE]{Parameter Estimation}
\acro{Bilby}[Bilby]{Bayesian Inference Library for \acs{CBC} \acs{GW} signal}
\acro{RAVEN}[RAVEN]{Rapid, on-source VOEvent Coincident Monitor}
\end{acronym}

\begin{document}

\title{Low-latency gravitational wave alert products and their performance at the time of the fourth LIGO-Virgo-KAGRA observing run}

\author[0000-0003-1314-4241]{Sushant Sharma Chaudhary*}
\affiliation{Institute of Multi-messenger Astrophysics and Cosmology, Missouri University of Science and Technology, Physics Building, 1315 N. Pine St., Rolla, MO 65409, USA}
\email{*sscwrk@mst.edu}

\author[0009-0008-9546-2035]{Andrew Toivonen$^{\dagger}$}
\affiliation{School of Physics and Astronomy, University of Minnesota, Minneapolis, Minnesota 55455, USA}
\email{$^{\dagger}$toivo032@umn.edu}

\collaboration{2}{These authors contributed equally to this work}

\author[0000-0003-3630-9440]{Gaurav Waratkar}
\affiliation{Department of Physics, IIT Bombay, Powai, Mumbai, 400076, India}

\author[0000-0001-6331-112X]{Geoffrey Mo}
\affiliation{LIGO Laboratory and Kavli Institute for Astrophysics and Space Research, Massachusetts Institute of Technology, 185 Albany Street, Cambridge, Massachusetts 02139, USA}

\author[0000-0003-0038-5468]{Deep Chatterjee}
\affiliation{LIGO Laboratory and Kavli Institute for Astrophysics and Space Research, Massachusetts Institute of Technology, 185 Albany Street, Cambridge, Massachusetts 02139, USA}

\author[0000-0002-7686-3334]{Sarah Antier}
\affiliation{Artemis, Observatoire de la C\^ote d'Azur, Universit\'e C\^ote d'Azur, Boulevard de l'Observatoire, 06304 Nice, France}

\author{Patrick Brockill}
\affiliation{Leonard E. Parker Center for Gravitation, Cosmology, and Astrophysics, University of Wisconsin-Milwaukee, Milwaukee, WI 53201, USA}

\author[0000-0002-8262-2924]{Michael W. Coughlin}
\affiliation{School of Physics and Astronomy, University of Minnesota, Minneapolis, Minnesota 55455, USA}

\author[0000-0001-8196-9267]{Reed Essick}
\affiliation{Canadian Institute for Theoretical Astrophysics, University of Toronto, Toronto, ON M5S 3H8} 
\affiliation{Department of Physics, University of Toronto, Toronto, ON M5S 1A7}
\affiliation{David A. Dunlap Department of Astronomy, University of Toronto, Toronto, ON M5S 3H4}

\author[0000-0001-9901-6253]{Shaon Ghosh}
\affiliation{Montclair State University, 1 Normal Ave. Montclair, NJ 07043}

\author[0000-0002-8445-6747]{Soichiro Morisaki}
\affiliation{Institute for Cosmic Ray Research, The University of Tokyo, 5-1-5 Kashiwanoha, Kashiwa, Chiba 277-8582, Japan}

\author[0000-0001-6308-211X]{Pratyusava Baral}
\affiliation{Leonard E. Parker Center for Gravitation, Cosmology, and Astrophysics, University of Wisconsin-Milwaukee, Milwaukee, WI 53201, USA}

\author[0000-0003-0918-0864]{Amanda Baylor}
\affiliation{Leonard E. Parker Center for Gravitation, Cosmology, and Astrophysics, University of Wisconsin-Milwaukee, Milwaukee, WI 53201, USA}

\collaboration{11}{}

\nocollaboration{41}

\author[0000-0002-4559-8427]{Naresh Adhikari}
\affiliation{Leonard E. Parker Center for Gravitation, Cosmology, and Astrophysics, University of Wisconsin-Milwaukee, Milwaukee, WI 53201, USA}

\author{Patrick Brady}
\affiliation{Leonard E. Parker Center for Gravitation, Cosmology, and Astrophysics, University of Wisconsin-Milwaukee, Milwaukee, WI 53201, USA}

\author[0000-0002-4289-3439]{Gareth Cabourn~Davies} 
\affiliation{University of Portsmouth, Portsmouth, PO1 3FX, United Kingdom}

\author[0000-0001-5078-9044]{Tito Dal Canton}
\affiliation{Universit\'e Paris-Saclay, CNRS/IN2P3, IJCLab, 91405 Orsay, France}

\author[0000-0002-3835-6729]{Marco Cavaglia}
\affiliation{Institute of Multi-messenger Astrophysics and Cosmology, Missouri University of Science and Technology, Physics Building, 1315 N. Pine St., Rolla, MO 65409, USA}

\author[0000-0003-3600-2406]{Jolien Creighton}  \affiliation{University of Wisconsin, Milwaukee, WI 53211, USA}

\author[0000-0003-0949-7298]{Sunil Choudhary}
\affiliation{Australian Research Council Centre of Excellence for Gravitational Wave Discovery (OzGrav), Australia}
\affiliation{Department of Physics, University of Western Australia, Crawley WA 6009, Australia}

\author[0000-0002-8661-4120]{Yu-Kuang Chu}
\affiliation{Leonard E. Parker Center for Gravitation, Cosmology, and Astrophysics, University of Wisconsin-Milwaukee, Milwaukee, WI 53201, USA}

\author{Patrick Clearwater}
\affiliation{Australian Research Council Centre of Excellence for Gravitational Wave Discovery (OzGrav), Australia}
\affiliation{Department of Physics, University of Western Australia, Crawley WA 6009, Australia}

\author{Luke Davis}
\affiliation{Australian Research Council Centre of Excellence for Gravitational Wave Discovery (OzGrav), Australia}
\affiliation{Department of Physics, University of Western Australia, Crawley WA 6009, Australia}

\author[0000-0003-1354-7809]{Thomas Dent}
\affiliation{IGFAE, Universidade de Santiago de Compostela, E-15782 Spain}

\author[0000-0002-3738-2431]{Marco Drago}
\affiliation{Università di Roma La Sapienza, I-00133 Roma, Italy and INFN, Sezione di Roma, I-00133 Roma, Italy}

\author[0000-0001-9178-5744]{Becca Ewing}
\affiliation{Department of Physics, The Pennsylvania State University, University Park, PA 16802, USA}
\affiliation{Institute for Gravitation and the Cosmos, The Pennsylvania State University, University Park, PA 16802, USA}

\author[0000-0002-7489-4751]{Patrick Godwin}
\affiliation{LIGO Laboratory, California Institute of Technology, Pasadena, CA 91125, USA}

\author[0000-0002-4320-4420]{Weichangfeng Guo}
\affiliation{Australian Research Council Centre of Excellence for Gravitational Wave Discovery (OzGrav), Australia}
\affiliation{Department of Physics, University of Western Australia, Crawley WA 6009, Australia}

\author{Chad Hanna}
\affiliation{Department of Physics, The Pennsylvania State University, University Park, PA 16802, USA}
\affiliation{Institute for Gravitation and the Cosmos, The Pennsylvania State University, University Park, PA 16802, USA}
\affiliation{Department of Astronomy and Astrophysics, The Pennsylvania State University, University Park, PA 16802, USA}
\affiliation{Institute for Computational and Data Sciences, The Pennsylvania State University, University Park, PA 16802, USA}

\author{Rachael Huxford}
\affiliation{Department of Physics, The Pennsylvania State University, University Park, PA 16802, USA}
\affiliation{Institute for Gravitation and the Cosmos, The Pennsylvania State University, University Park, PA 16802, USA}

\author[0000-0002-5304-9372]{Ian Harry}
\affiliation{University of Portsmouth, Portsmouth, PO1 3FX, United Kingdom}

\author{Erik Katsavounidis}
\affiliation{LIGO Laboratory and Kavli Institute for Astrophysics and Space Research, Massachusetts Institute of Technology, 185 Albany Street, Cambridge, Massachusetts 02139, USA}

\author[0000-0001-8143-9696]{Manoj Kovalam}
\affiliation{Australian Research Council Centre of Excellence for Gravitational Wave Discovery (OzGrav), Australia}
\affiliation{Department of Physics, University of Western Australia, Crawley WA 6009, Australia}

\author[0000-0001-6728-6523]{Alvin K.Y. Li}
\affiliation{LIGO Laboratory, California Institute of Technology, Pasadena, CA 91125, USA}

\author[0000-0001-9769-531X]{Ryan Magee}
\affiliation{LIGO Laboratory, California Institute of Technology, Pasadena, CA 91125, USA}

\author[0009-0000-4183-7876]{Ethan Marx}
\affiliation{LIGO Laboratory and Kavli Institute for Astrophysics and Space Research, Massachusetts Institute of Technology, 185 Albany Street, Cambridge, Massachusetts 02139, USA}

\author{Duncan Meacher}
\affiliation{Leonard E. Parker Center for Gravitation, Cosmology, and Astrophysics, University of Wisconsin-Milwaukee, Milwaukee, WI 53201, USA}

\author[0000-0002-8230-3309]{Cody Messick}
\affiliation{Leonard E. Parker Center for Gravitation, Cosmology, and Astrophysics, University of Wisconsin-Milwaukee, Milwaukee, WI 53201, USA}

\author[0000-0001-6540-5517]{Xan Morice-Atkinson} \affiliation{University of Portsmouth, Portsmouth, PO1 3FX, United Kingdom}

\author[0009-0003-4044-0334]{Alexander Pace}
\affiliation{Department of Physics, Pennsylvania State University, University Park, PA 16802, USA}

\author[0000-0003-1556-8304]{Roberto De Pietri}
\affiliation{Dipartimento di Scienze Matematiche, Fisiche e Informatiche,
Università di Parma, Parco Area delle Scienze 7/A,I-43124 Parma, Italy}
\affiliation{INFN, Sezione di Milano Bicocca, Gruppo Collegato di Parma, I-43124 Parma, Italy}

\author{Brandon Piotrzkowski}
\affiliation{Leonard E. Parker Center for Gravitation, Cosmology, and Astrophysics, University of Wisconsin-Milwaukee, Milwaukee, WI 53201, USA}

\author[0000-0003-2147-5411]{Soumen Roy}
\affiliation{Nikhef, Science Park 105, 1098 XG Amsterdam, The Netherlands}
\affiliation{Institute for Gravitational and Subatomic Physics (GRASP), Utrecht University, Princetonplein 1, 3584 CC Utrecht, The Netherlands}

\author{Surabhi Sachdev}
\affiliation{Leonard E. Parker Center for Gravitation, Cosmology, and Astrophysics, University of Wisconsin-Milwaukee, Milwaukee, WI 53201, USA}
\affiliation{School of Physics, Georgia Institute of Technology, Atlanta, GW 30332, USA}

\author[0000-0001-9898-5597]{Leo P. Singer}
\affiliation{Astrophysics Science Division, NASA Goddard Space Flight Center, Code 661, Greenbelt, MD 20771, USA}
\affiliation{Joint Space-Science Institute, University of Maryland, College Park, MD 20742, USA}

\author{Divya Singh}
\affiliation{Department of Physics, The Pennsylvania State University, University Park, PA 16802, USA}
\affiliation{Institute for Gravitation and the Cosmos, The Pennsylvania State University, University Park, PA 16802, USA}

\author[0000-0002-6167-6149]{Marek Szczepanczyk}
\affiliation{Department of Physics, University of Florida, Gainesville, FL 32611-8440, USA}

\author{Daniel Tang}
\affiliation{Australian Research Council Centre of Excellence for Gravitational Wave Discovery (OzGrav), Australia}
\affiliation{Department of Physics, University of Western Australia, Crawley WA 6009, Australia}

\author[0000-0002-2728-9508]{Max Trevor}
\affiliation{University of Maryland, College Park, MD 20742, USA}

\author{Leo Tsukada}
\affiliation{Department of Physics, The Pennsylvania State University, University Park, PA 16802, USA}
\affiliation{Institute for Gravitation and the Cosmos, The Pennsylvania State University, University Park, PA 16802, USA}

\author[0000-0001-7983-1963]{Verónica Villa-Ortega}
\affiliation{IGFAE, Universidade de Santiago de Compostela, E-15782 Spain}

\author[0000-0001-7987-295X]{Linqing Wen}
\affiliation{Australian Research Council Centre of Excellence for Gravitational Wave Discovery (OzGrav), Australia}
\affiliation{Department of Physics, University of Western Australia, Crawley WA 6009, Australia}

\author[0000-0001-9138-4078]{Daniel Wysocki}
\affiliation{Leonard E. Parker Center for Gravitation, Cosmology, and Astrophysics, University of Wisconsin-Milwaukee, Milwaukee, WI 53201, USA}

\newcommand{\msun}{\ensuremath{M_{\odot}}}
\newcommand{\mc}{\ensuremath{\mathcal{M}_c}}
\newcommand{\hubbleunit}{\ensuremath{\text{km s}^{-1}~\text{Mpc}^{-1}}}
\newcommand{\totalinjections}{\ensuremath{5\times 10^4}}
\newcommand{\BNSrec}{\ensuremath{1489}}
\newcommand{\NSBHrec}{\ensuremath{1105}}
\newcommand{\BBHrec}{\ensuremath{1920}}

\newcommand{\BNSinj}{\ensuremath{40.9\%}}
\newcommand{\NSBHinj}{\ensuremath{35.8\%}}
\newcommand{\BBHinj}{\ensuremath{23.3\%}}

\newcommand{\hasns}{\ensuremath{\texttt{HasNS}}}
\newcommand{\hasremnant}{\ensuremath{\texttt{HasRemnant}}}
\newcommand{\hasmassgap}{\ensuremath{\texttt{HasMassGap}}}

\newcommand{\tsuperevent}{t_{\text{superevent}}}
\newcommand{\tmerger}{t_{\text{0}}}
\newcommand{\tevent}{t_{\text{event}}}
\newcommand{\tadvreq}{t_{\text{ADV\_REQ}}}
\newcommand{\tgcn}{t_{\text{GCN\_PRELIM}}}
\newcommand{\temcoinc}{t_{\text{EM\_COINC}}}
\newcommand{\travenalert}{t_{\text{RAVEN\_ALERT}}}
\newcommand{\gwcelery}{\texttt{GWCelery}}
\newcommand{\celery}{\texttt{Celery}}
\newcommand{\flask}{\texttt{Flask}}
\newcommand{\redis}{\texttt{Redis}}

\newcommand{\publicalertthresholdbefortrials}{\ensuremath{ \leq 1.6 \times 10^{-4}}\,Hz (fourteen per day)}
\newcommand{\publicalertthresholdaftertrails}{\ensuremath{ \leq 2.3 \times 10^{-5}}\,Hz (two per day)}
\newcommand{\significantalertthresholdbeforetrials}{\ensuremath{\leq 3.9 \times 10^{-7}}\,Hz (one per month)}
\newcommand{\significantalertaftertrialsmdc}{\ensuremath{\leq 6.4 \times 10^{-8}}\,Hz (one per 6 months)}
\newcommand{\significantalertthresholdaftertrials}{\ensuremath{\leq 7.7 \times 10^{-8}}\,Hz (one per 5 months)}
\newcommand{\burstalertthresholdbeforetrials}{\ensuremath{\leq 3.2 \times 10^{-8}}\,Hz (one per year)}
\newcommand{\burstalertthresholdaftertrials}{\ensuremath{\leq 7.9 \times 10^{-9}}\,Hz (one per 4 year)}

\newcommand{\dc}[1]{{\textcolor{red}{{[DC: {#1}]}} }}







\begin{abstract}
Multi-messenger searches for \ac{BNS} and \ac{NSBH} mergers are currently one of the most exciting areas of astronomy. The search for joint electromagnetic and neutrino counterparts to \ac{GW}s has resumed with \ac{O4}. To support this effort, public semi-automated data products are sent in near real-time and include localization and source properties to guide complementary observations. In preparation for \ac{O4}, we have conducted a study using a simulated population of compact binaries and a \ac{MDC} in the form of a real-time replay to optimize and profile the software infrastructure and scientific deliverables. End-to-end performance was tested, including data ingestion, running online search pipelines, performing annotations, and issuing alerts to the astrophysics community. We present an overview of the low-latency infrastructure and the performance of the data products that are now being released during \ac{O4} based on the \acs{MDC}. We report the expected median latency for the preliminary alert of full bandwidth searches (29.5\,s) and show consistency and accuracy of released data products using the \ac{MDC}.  For the first time, we report the expected median latency for triggers from early warning searches (-3.1\,s), which are new in \ac{O4} and target neutron star mergers during inspiral phase. This paper provides a performance overview for \acs{LVK} low-latency alert infrastructure and data products using the \ac{MDC} and serves as a useful reference for the interpretation of \ac{O4} detections. 

\end{abstract}




\section{Introduction} \label{sec:intro}

As of May 24 2023\footnote{\url{https://observing.docs.ligo.org/plan}}, \acf{O4} is underway, following a series of observing runs, which have reported the detection of the first \ac{BBH} in \ac{O1} \citep{AbEA2016b}, the detection of the first \ac{BNS} merger \citep{AbEA2017b} and associated electromagnetic counterparts AT2017gfo \citep{CoFo2017,SmCh2017,AbEA2017f} and GRB170817A~\citep{GoVe2017,SaFe2017,AbEA2017e} in the \ac{O2}, and \ac{NSBH} \citep{AbEA2021} in the \ac{O3}. 
Focusing on \ac{NS} mergers, there are a variety of science cases for their multi-messenger counterpart searches and detections, including measurements of the \ac{NS} \ac{EoS} \citep{BaJu2017, MaMe2017, CoDi2018b, CoDi2018, CoDi2019b, AnEe2018, MoWe2018,RaPe2018,Lai2019,DiCo2020,Huth:2021bsp}, the Hubble constant \citep{CoDi2019,CoAn2020,2017Natur.551...85A,HoNa2018,DiCo2020}, and $r$-process nucleosynthesis \citep{ChBe2017,2017Sci...358.1556C, CoBe2017,PiDa2017,RoFe2017,SmCh2017,WaHa2019,KaKa2019}. 

\begin{figure*}[ht!]
\plotone{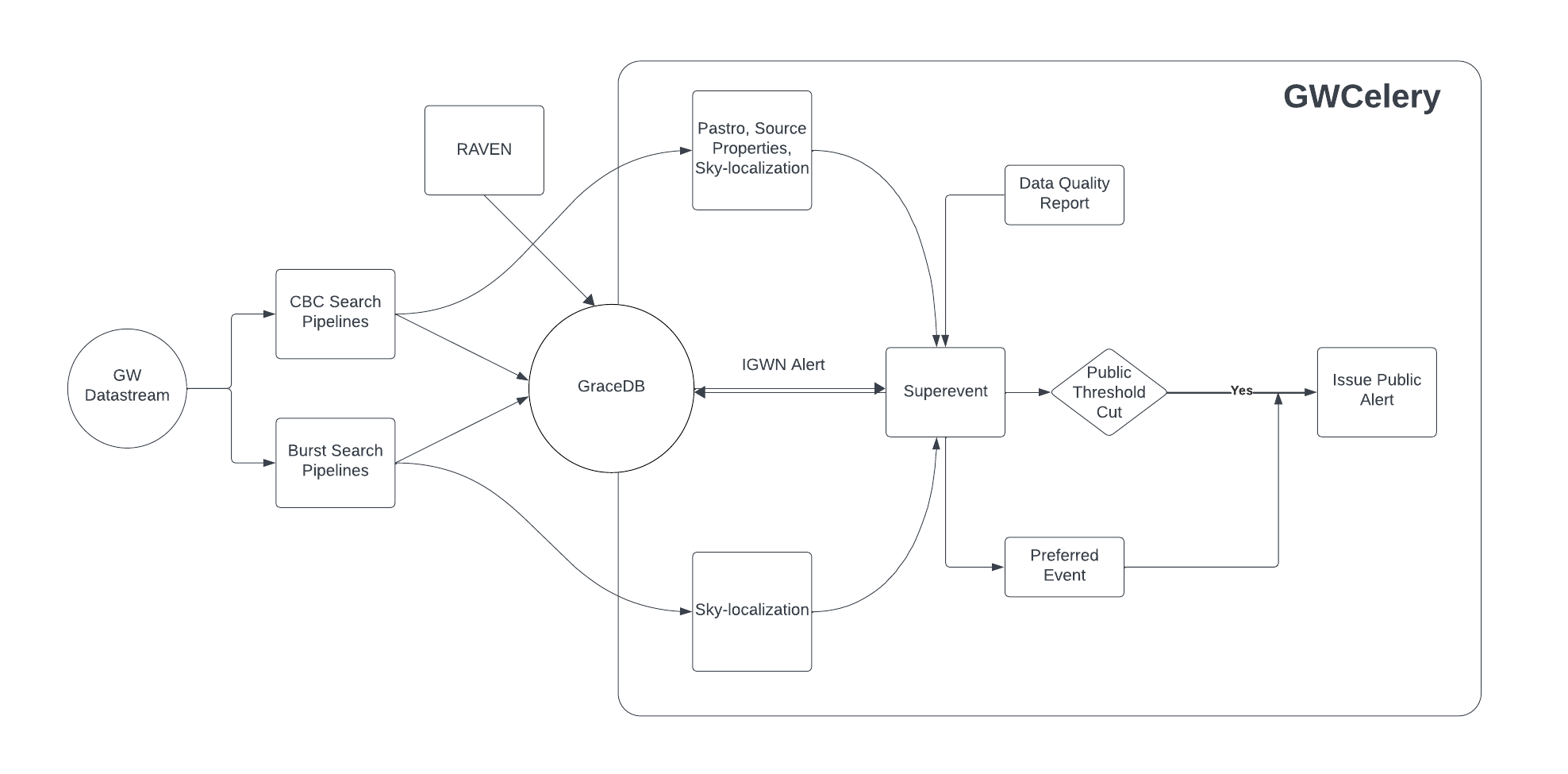}
\caption{Task flow of low-latency alert infrastructure. The process begins with search pipeline trigger(s) on a candidate in the GW datastream, which are passed through data quality checks and compiled into a superevent. If the preferred event from the superevent passed the significant FAR cut, a preliminary alert is sent out to the public.}
\label{fig:flow}
\end{figure*}

The \ac{LVK}'s real time alert infrastructure depends on several components. Broadly this includes low-latency data calibration and transfer, running of modeled and unmodeled online searches (see Section~\ref{sec:searches} for a brief description), and maintaining the state of events in \ac{GraceDB} following the discovery. In addition to \ac{GraceDB},\footnote{\url{https://gracedb.ligo.org/}} which serves as both the database and as an internal and external web view, the alert infrastructure includes \texttt{igwn-alert},\footnote{\url{https://igwn-alert.readthedocs.io}} an internal messaging system to communicate the state of events, and \gwcelery,
\footnote{\url{https://git.ligo.org/emfollow/gwcelery}}
\footnote{\url{https://rtd.igwn.org/projects/gwcelery/en/latest/}} a task queue, to cluster, annotate, and orchestrate the events, as well as publish public alerts\footnote{For example, those hosted by SCiMMA and NASA} for the community to subscribe to. Figure~\ref{fig:flow} shows the task flow of the \ac{LLAI} for candidate events. The current \ac{LLAI} is a significantly upgraded version of the infrastructure used earlier, described in \citealt{AbEA2019b}, and used in the more recently reported early-warning system, reported in \citealt{MaCh2021}. The primary changes compared to previous observing run are the transition from a XMPP based pubsub system internally to
Kafka-based messaging provided by the SCiMMA broker, removal of timeouts at various places of synchronization
and instead relying on labels on \ac{GraceDB} to keep track of the state of the superevent, and reconfiguring the snapshotting
configuration for {\redis} database to strike a balance between fault tolerance and avoid filling disk quota. Aside from
this, upgrades to software versions of the dependencies and running computationally expensive resources on specific
pool of modern hardware contributed to a improvement compared to previous observing run. The \acs{LVK} Alert User Guide\,\footnote{\url{https://emfollow.docs.ligo.org/userguide/}} constitutes a living document where information and updates about this system are regularly communicated to the broader community. Further discussion of \ac{GraceDB}, \texttt{igwn-alert}, and \gwcelery \ is provided in the supplementary material.


To prepare for \acs{O4} and demonstrate performance across a variety of software and alert system improvements, we carried out a \acf{MDC}. This \acs{MDC} constitutes a testing environment for the \acs{LLAI} to prepare for O4, producing repeated sets of 40 days of data from O3, with associated simulations of \acp{CBC} to stress test the system. While the rates of simulated events (see Section~\ref{sec:results} for a description of the data set) were much higher than that expected for O4, this high rate was designed to test the various components of searches, the alert system and the scientific deliverables before heading into O4; these include, for example, tests of the detection efficiency of online real time low-latency searches, the rapid estimation of the binary system properties, and their associated sky localizations. 

In this paper, we describe the details of the alert system for \acs{O4} and its performance based on this \ac{MDC}.
In addition, we provide an overview of the detection performance of real time searches, along with the consistency and accuracy of alert data products.
Section~\ref{sec:pipeline} provides an overview of the LLAI and the scientific data products reported, while
Section~\ref{sec:mdc} describes the properties of the MDC, including the motivations for the choices made.
Section~\ref{sec:results} reports the properties of the LLAI as of the beginning of O4, as measured by the MDC, and
Section~\ref{sec:conclusion} describes the conclusions and prospects for future development.

\section{Overview of alert system and scientific products}
\label{sec:pipeline}

\subsection{Searches}
\label{sec:searches}

Low-latency \ac{GW} searches consist of two categories: ``modeled'' \ac{CBC} \citep{AbEA2021d} and ``unmodeled'' (Burst) \citep{AbEA2021b} searches. Modeled \ac{CBC} searches target \ac{BNS}, \ac{NSBH}, or \ac{BBH} \citep{AbEA2021d}; unmodeled searches look for signals with generic morphologies from a wide variety of astrophysical sources like core-collapse of massive stars, magnetar star-quakes, and other sources, in addition to compact binary mergers \citep{AbEA2021b,AbEA2021c}. For the purpose of this \ac{MDC} analysis, we focus on \ac{CBC} searches, but also report latencies of injections found by Burst pipelines. \ac{CBC} searches can be categorized as early warning, referring to pre-merger searches \citep{Sachdev_2020, kovalam_2022}, or full bandwidth, referring to post-merger, based on how the search truncates their templates. Each search produces candidate \ac{GW} triggers and assigns them ranking statistic values and \acp{FAR}; the \ac{FAR} for a trigger in a given search pipeline is defined as the expected rate of triggers due to detector noise, in that pipeline, with equal or higher ranking. Each search pipeline has different and independent methods of generating and ranking triggers and estimating the noise background and thus the \acs{FAR}; 
for details see~\cite{Messick:2016aqy, Aubin:2020goo, s_hooper_2012,j_luan_2012, Usman_2016, DalCanton:2020vpm, piotrzkowski2022searching}. In addition, the probability of astrophysical origin, \pastro, for a trigger is calculated for \ac{CBC} searches, which is described in detail in Section \ref{sec:alerts}. In the following, we briefly describe key aspects of each pipeline participating in low-latency searches. 

\subsubsection{GstLAL}

\ac{GstLAL} is a stream-based matched filtering algorithm capable of detecting \ac{GW} signals within seconds of their arrival on Earth~\citep{Messick:2016aqy, Tsukada:2023edh, Ewing:2023qqe}. \ac{GstLAL} uses a template bank of $\sim10^6$ \ac{CBC} waveforms in order to filter the full \ac{BNS}, \ac{NSBH}, and \ac{BBH} regions of the parameter space~\citep{Sakon:2022ibh}. The template bank is divided into $\sim 10^3$ bins of time-sliced \ac{SVD} waveforms according to the \ac{LLOID} method~\citep{Cannon:2011vi}. These waveforms are used to filter the strain data producing an output \ac{SNR} timeseries.  Peaks in the \ac{SNR} time-series which pass a threshold of $4.0$ are stored as ``triggers”. These form candidates which may be coincident among two or more detectors or observed in only a single detector. Significance is assigned to each candidate using the likelihood ratio ranking statistic which is then mapped to a false alarm probability and corresponding \ac{FAR}~\citep{Cannon:2015gha}. Candidates are finally uploaded to \ac{GraceDB} after aggregating them across SVD bins by maximum \ac{SNR}. \ac{GstLAL} carries out both an early warning search, and a full bandwidth search. The early warning search targets low redshift \ac{BNS} events that can be detected $\sim$ 10 - 60 s before merger, using templates with non-spinning component masses between 0.95 \msun \ and 2.4 \msun \ \citep{Sachdev_2020}. The full bandwidth search covers the entire \ac{CBC} template bank parameter space.

\ac{GstLAL} uses the multi-component \acs{FGMC} method for assigning a probability of astrophysical origin to candidates~\citep{Kapadia:2019uut,PhysRevD.91.023005}. The probability that the signal originates from each \ac{CBC} source category is also assigned. Triggers from each category (\ac{BNS}, \ac{NSBH}, \ac{BBH} and terrestrial) are treated as realizations of independent Poisson processes. The rate of detectable triggers characterizing the Poisson process corresponding to each foreground category is approximated from the astrophysical rate estimates yielded by offline FGMC analyses of past observing runs while accounting for the change in sensitive spacetime volume between the past and ongoing runs. Misclassification among astrophysical source categories is accounted for by computing the probability of migration between injected and recovered templates across the entire bank semi-analytically under the Gaussian noise approximation~\citep{Fong:2018elx}. With the rates and migration probabilities precomputed, \pastro is estimated in low-latency from trigger data comprising the likelihood ratio ranking statistic assigned to said trigger and the matched template.

\subsubsection{MBTA}

The \ac{MBTA} pipeline carries out an early warning and main, full bandwidth search and performs matched filtering per frequency band to reduce computational costs \citep{Aubin:2020goo}.
The main instance of the pipeline is searching for binaries with total masses ranging from 2 to 500 \msun \ and mass ratio smaller than 50. 

The pipeline includes signal-consistency checks to help distinguish astrophysical signals from background.
As for the O3 offline analysis \citep{Andres:2021vew} the probability of astrophysical origin of \ac{GW} candidate events is derived from the expected rate of astrophysical events and background candidates at the recovered chirp mass, mass ratio and ranking statistic.
The foreground distribution has been estimated by performing injections of simulated \ac{BNS}, \ac{BBH} and \ac{NSBH} signals into LIGO-Virgo O3a data which are then analyzed by the \ac{MBTA} pipeline.
This method computes \pastro and source classification.
It has been extended to also provide EM-Bright information to determine the likelihood of an electromagnetic counterpart, as covered in Section \ref{sec:alerts}.


\subsubsection{PyCBC Live}

\pcl is a matched filtering pipeline designed to detect \ac{CBC} events by comparing the incoming \ac{GW} signal to a template bank of waveforms \citep{Nitz:2018rgo,DalCanton:2020vpm}. 
Two \pcl searches were employed; one is a full bandwidth search for a wide range of signals, the other is an early-warning configuration for which the templates are truncated at certain frequencies before merger~\citep{Nitz:2020vym}. The full bandwidth template bank contains 412,575 templates, covering total masses from 2 to 500\,\msun, and mass ratios from 1 to 100~\citep{Roy:2017qgg, Roy:2017oul}. 
The early-warning template bank contains $\sim$$4700$ templates with component masses in the range 1 to 3\,\msun, truncated at a set of frequencies designed to give early warnings at regular intervals before the merger. 

The matched filtering algorithm produces a time series of \ac{SNR} values, and only triggers with \ac{SNR} $\geq 4.5$ are considered for further analysis.
The \ac{SNR} is then re-weighted according to signal-consistency tests in each detector, and using multi-detector properties determined by the distribution of source extrinsic parameters (time difference, phase difference and amplitude ratios) for the signal population~\citep{Nitz:2017svb}.

In order to assess the frequency of a coincident noise signal which would be ranked greater or equal to a given detection, \pcl assigns a FAR value by comparing the candidate to time-shifted background from the last several hours~\citep{Nitz:2018rgo}. 
The FAR values for injections recovered during the MDC are subject to a substantial upward bias due to the high rate of high-\ac{SNR} injected events, which significantly influences the background estimation. 

For single-detector candidates, if strict criteria on signal consistency tests are passed, a FAR is assigned by comparing the candidate's re-weighted \ac{SNR} to the noise trigger distribution via a template-dependent exponential fit~\citep{DalCanton:2020vpm}.
The exponential fit is performed using the original data without injections, thus the FAR calculation for single-detector events is not subject to contamination from injections.  
Single-detector candidate signals are only considered for potentially electromagnetically bright signals, as \ac{BBH} signals in low-latency are unlikely to yield multimessenger counterparts, and higher-mass templates are more susceptible to glitch contamination due to their shorter duration.
Single-detector early-warning candidate signals are also not considered, as the poor localization of single-detector events is not of interest for pre-merger alerts.

For full bandwidth events, PyCBC Live also calculates the probability of astrophysical origin \pastro, based on the FAR value, the trigger \ac{SNR}, the approximate distributions of signal and noise events over template chirp mass, and the sensitivities of observing detectors~\citep{pycbc_pastro_o4}.  This \pastro is then combined with estimates of the relative probabilities of different source classes, also based on template chirp mass; for details see~\cite{villaortega2022astrophysical}.

In the later part of O3, an additional step was performed to optimize event \ac{SNR} over template masses and spins~\citep{DalCanton:2020vpm}.  Our MDC results include additional events produced via this optimization; however, it was removed from the search configuration deployed at the start of O4 in order to reduce complexity and computational load.  We do not expect major differences in the search outputs detailed here due to the change.

\subsubsection{SPIIR}

The \ac{SPIIR} pipeline is designed to achieve lower delays in signal detection and differs from other pipelines in multiple aspects. As the name suggests, \ac{SPIIR} uses a time domain counterpart of matched filtering~\citep{s_hooper_2012,j_luan_2012} as its primary filter. This method breaks down millions of \ac{CBC} templates into a few hundred thousand \acs{IIR} filters to perform match filtering in the time domain, further accelerated by the use of GPUs. The \ac{SPIIR} pipeline implements a computational coherent network search approach~\citep{Bose:1999pj,harry_2011} to select the \ac{GW} candidate events with low-latency, achieved with the help of \acs{SVD}~\citep{wen_2008}. The pipeline performs a full bandwidth search for \ac{BBH}, \ac{BNS}, and \ac{NSBH} sources, and an early warning search for \ac{BNS} and \ac{NSBH} sources. \ac{SPIIR} has demonstrated its performance in past LIGO-Virgo runs~\citep{chu2021spiir, kovalam_2022}.

\ac{SPIIR} is introducing a new two-step \pastro calculation for O4. In the first step, the pipeline calculates the two-component \pastro of the trigger based on the FGMC two-component method by~\citet{PhysRevD.91.023005} and \citet{Kapadia:2019uut}. This assigns the probability of the trigger's astrophysical or terrestrial origin. In the second step, it further classifies the probability of astrophysical origin into \ac{NSBH}, \ac{BBH}, and \ac{BNS}, based on the chirp mass method~\citep{villaortega2022astrophysical}.

\subsubsection{Coherent WaveBurst}

\ac{cWB} \citep{klimenko2008coherent, Klimenko:2015ypf, drago2021coherent} is an excess power algorithm using minimal assumptions on the \ac{GW} signature. cWB decomposes the GW strain data using a wavelet transform (\cite{Necula:2012zz}). It then selects coherent signal power in multiple detectors and applies a maximum likelihood approach to select GW events. The calculation of the likelihood over the sky allows for building a sky map that characterizes the probability of the GW source sky location. A new feature with respect to O3 cWB analyses has been implemented for the significance assessment - a machine learning algorithm based on XGBoost~\citep{Mishra:2022ott, Szczepanczyk:2022urr}. In low-latency, cWB analyzes 180~s data segments overlapping every 30~s. The alerts are created up to a latency of around 1~minute.

\subsubsection{oLIB}
The omicron-LALInferenceBurst (oLIB) pipeline is a short duration 
($\lesssim$ 1 second) unmodeled detection pipeline that is sensitive to a wide variety of sources that includes, but is not limited to, CBCs \citep{Lynch_2017}. As such, oLIB makes very minimal assumptions about the astrophysical source type of the emission. The search is performed hierarchically. First, data from individual interferometers  is analyzed with the Omicron trigger generator algorithm \citep{Robinet_2020}. Omicron identifies regions in the time-frequency plane of excess power. Triggers that are coincident in time and frequency between interferometers are then followed up with a coherent Bayesian analysis using LIB. LIB models the data with a single sine-Gaussian wavelet, calculating two Bayes factors. Each of these Bayes factors is expressed as the natural logarithm of the evidence ratio of two hypotheses: (1) a GW signal versus Gaussian noise (BSN) and (2) a coherent GW signal versus incoherent noise transients (BCI). Ultimately, these two Bayes factors are used to construct a likelihood ratio $\Lambda$ that is used as the final search statistic. 

\subsubsection{RAVEN}
\ac{RAVEN} is a multi-messenger pipeline that searches for coincidence between GW candidates and other astronomical detections, such as gamma-ray bursts (GRBs) and neutrino bursts \citep{urban2016, cho2019low, piotrzkowski2022searching}. RAVEN ingests events submitted to the \ac{GCN}\footnote{\url{https://gcn.nasa.gov/}} (\citealt{SiRa2023}) into GraceDB, queries GraceDB to look for a corresponding GW candidate, and calculates the joint false alarm rate to determine whether to send a public alert.

\subsection{Low-Latency Alert Infrastructure}

\subsubsection{GraceDB}
\ac{GraceDB}\footnote{\url{https://gracedb.ligo.org/}} is the central location that houses \ac{GW} event candidates and analyses for transient searches. \ac{GW} event candidate data in \ac{GraceDB} can be viewed and manipulated on the web or through the use of a RESTful API. A permission structure exists to show only proprietary data to LVK users versus data that is available to the general public. State changes in \ac{GraceDB} (which may take the form of new event uploads/annotations, new superevent uploads/annotations, log/file updates, etc.) are communicated to LVK and external users and processes via the \texttt{igwn-alert} system.
 
At its core, \ac{GraceDB} is, architecturally, a standard Web/API application. \ac{GraceDB} is hosted in a high-availability configuration in Amazon AWS. A PostgreSQL backend is powered by a Django web framework. External requests are served by Apache acting as a reverse-proxy for a Gunicorn-based WSGI HTTP server. Files are stored on an NFS (Amazon EFS) filesystem, and low-latency analyses stream data from the detectors and upload candidate events to \ac{GraceDB} via a representational state transfer (REST) API.

\subsubsection{igwn-alert}
 
\texttt{igwn-alert}\footnote{\url{https://igwn-alert.readthedocs.io}} is an alert data stream based on kafka and leverages SCiMMA (Scalable Cyberinfrastructure for Multimessenger Astronomy) infrastructure for data delivery. Client-side tools maintained by IGWN Computing and Software (CompSoft) allow users to listen and respond to \texttt{igwn-alert} messages. \texttt{igwn-alert} messages are machine-readable (JSON), so as to be read by automated followup processes. \texttt{igwn-alert} listeners act on notifications from \ac{GraceDB} and are used to launch follow-up analyses (e.g., Superevent creation, parameter estimation, sky localization, etc.). Results from follow-up analyses are then uploaded and stored in \ac{GraceDB}. \ac{GraceDB} and \texttt{igwn-alert} are the orchestrator and source-of-truth for external observers and follow-up processes.

\subsubsection{GWCelery}

\gwcelery\footnote{\url{https://git.ligo.org/emfollow/gwcelery}} is a distributed task queue application for orchestrating and annotating GW alerts.
At its core, it is a {\celery} \footnote{\url{https://docs.celeryq.dev/}} application. Some of the advantages of a distributed task
queue, like {\celery}, include handling of asynchronous tasks, easy scalability based on requirement,
designing \emph{canvas} workflows, setting conditions to retry individual parts of a canvas, easy error
handling, and running periodic tasks. {\celery} is fault tolerant and preserves the state of tasks
in a result backend, and communicates with it using a messaging broker. {\gwcelery} uses {\redis} as
both a broker and a backend for the application. {\gwcelery} also contains a {\flask} application
to provide a web interface to run routine tasks which require human interaction with the application,
or to handle situations where part of a canvas are to be executed manually, overriding the
automated processing. We run all {\gwcelery} processes using HTCondor, used for job
scheduling in the LIGO Data Grid. The major subsystems of {\gwcelery} are:
\begin{itemize}
    \item The listener for IGWN alerts, which is pubsub system that \ac{GraceDB} uses to push machine-readable notifications about its state.
    \item The Superevent Manager, which clusters individual GW candidates into \emph{superevents}\footnote{\url{https://gwcelery.readthedocs.io/en/latest/gwcelery.tasks.superevents.html\#gwcelery-tasks-superevents-module}}.
    \item The External Trigger Manager which listens for and correlates candidates from external facilities to spot coincidences with GW events.
    \item The GCN and SCiMMA alert producer that disseminates GW candidate information for external consumption.
    \item The Orchestrator, which executes the per-event annotation workflow. This involves having the data products ready for sending alerts for the superevent, and broadly includes computing rapid sky-localization and source properties for individual events and updating the state of the superevent, and launching parameter estimation runs.
\end{itemize}

\subsection{Selection of the public GW event candidate} \label{significant}

When a potential \ac{GW} signal appears in the detector, the low-latency search pipelines analyze the signal and produce event candidates. Each pipeline can report multiple candidates for a single \ac{GW} signal. The event candidates, reported within a specific time window (1\,s around coalescence time for \ac{CBC} searches, and 1\,s around trigger time for burst searches), are collected and grouped as a \textit{superevent}. In the collection, one \ac{GW} event candidate is identified as the preferred event (defined as the event with the highest network \ac{SNR} for \ac{CBC} pipelines whereas lowest FAR for burst pipelines), and its properties and by-products are prepared for release to the public, which include the merger time, FAR, sky localization, and classification (see Section~\ref{sec:alerts}). Alerts are released publicly when a FAR passes the public alert threshold, currently FAR \publicalertthresholdbefortrials. An alert is labeled as \textit{significant} when a \ac{CBC} alert passes a FAR threshold of FAR \significantalertthresholdbeforetrials \ or when an unmodeled burst alert passes a FAR threshold of FAR \burstalertthresholdbeforetrials. Since multiple \ac{CBC} and burst searches run in low-latency, to account for the trials factor from these different searches with statistically independent false alarms, events from \ac{CBC} searches are labeled significant when a FAR passes a threshold of FAR \significantalertthresholdaftertrials, whereas events from burst target searches require a FAR \burstalertthresholdaftertrials. In the MDC, however, \ac{CBC} trials factor of 6 was used due to which \ac{CBC} events were labeled significant when a FAR passes a threshold of FAR \significantalertaftertrialsmdc.  Considering the trials factor accounting for all searches, the public alert threshold is FAR \publicalertthresholdaftertrails. Alerts that meet the public alert threshold but not the significant threshold are labeled \textit{low significance}. RAVEN only uses the significant FAR thresholds when assessing its joint FAR for publication, with additional trials factors to compensate for listening to multiple GW pipelines. The \acs{LVK} Alert User Guide\footnote{\url{https://emfollow.docs.ligo.org/userguide/}} should be referenced for up-to-date information on trials factors during observing runs.

\subsection{Alert Contents}
\label{sec:alerts}

Public alerts are sent in order to inform the greater astronomical community of \ac{GW} events to enable multi-messenger follow-up of these events. These alerts are distributed both via GCN and the Scalable Cyberinfrastructure to support Multi-Messenger Astrophysics\footnote{\url{https://scimma.org/}} (SCiMMA) project in order to reach maximum consumers through two broad bases of subscribers. The alerts come in two types: notices, which are machine-readable and come in a variety of formats, and GCN circulars, which are human-readable. 

There are five types of notices that may be sent out for a candidate event: Early Warning, Preliminary, Initial, Update, and Retraction. Early Warning Notices arise from dedicated pre-merger search pipelines, potentially enabling the release of alerts seconds before merger \citep{MaCh2021}. A first Preliminary Notice is sent out when an event candidate of a superevent exceeds the public FAR threshold. Following a timeout, the preferred event is determined and a second Preliminary Notice is then issued (even if the preferred event candidate remains unchanged). Both Early Warning and Preliminary Notices are sent out if the candidate passes automatic data quality checks \citep{Arnaud:2022iiy}. These data quality checks are carried out by the Data Quality Report framework\footnote{\url{https://docs.ligo.org/detchar/data-quality-report/}}, and include checks for terrestrial noise and stationarity of the data, among others. In certain cases, such as when manual data quality checks yield suspicions on the astrophysical nature of the candidate, a Retraction Notice may be sent. If, however, the Early Warning or Preliminary Notice passes human vetting, then an Initial Notice is sent out accompanied by a GCN Circular to announce the detection. The final type of notice, an Update Notice, is used to send out improved estimates of alert contents based on parameter estimation if they become available. Included in each alert is an estimate of the event's probability of astrophysical origin, or \pastro. This is broken up into four categories that sum to 1 by definition: P(BNS), P(NSBH), P(BBH), and P(Terrestrial), where the mass boundary between \ac{NS} and \ac{BH} is set at 3\,$M_\odot$. If the superevent is coincident with a GCN candidate, the various data products concerning the joint candidate are included, such as the time delay, joint FAR \footnote{\url{https://ligo-raven.readthedocs.io}}, and combined sky map if applicable.

\paragraph{\textbf{Sky localization}}

One of the key data products to enable multi-messenger follow-up is the rapid inference of the sky localization from \ac{GW} observations. This sky localization consists of the posterior probability distribution of the source location in the sky. 
The sky localization, mapped either over a 2D map of right ascension and declination, or a 3D volume which also includes a distance estimate, is known as a ``sky map.''
Sky localization (and parameter estimation more generally) is conducted in multiple stages once a candidate is identified.

For CBC sources, \ac{BAYESTAR},
a rapid sky localization algorithm \citep{2016PhRvD..93b4013S}, is used to generate sky maps, and may be updated by \texttt{Bilby} (Section~\ref{PE}), a python-based parameter estimation pipeline that uses stochastic sampling methods \citep{2019ApJS..241...27A, Romero-Shaw:2020owr}. Sky maps from \acs{BAYESTAR} are released with Preliminary Notices and sky maps from \texttt{Bilby} are released in Update Notices. Additionally, \ac{cWB} also generates localizations \citep{2011PhRvD..83j2001K}.

The sky map is stored as \acs{FITS} file using the \ac{HEALPix} \citep{GoHi2005} framework in the \ac{MOC} representation \citep{FeBo2014}; flattened versions at a fixed HEALPix grid size are also available for superevents. MOC sky maps use adaptive division of the HEALPix grid, focusing areas of highest resolution on regions of highest probability with minimal information loss. Sky maps are made available both through the distributed alert as well as uploaded on \ac{GraceDB}, where they are available for direct download. If there is a coincidence with a GRB candidate that has a sky localization, we compute the overlap integral with the GW sky map information. This weighted sky map is then included in the alert. The technique of combining sky maps from two independent datasets is laid out for the first time in \citep{urban2016}, under the signal hypothesis of the Bayesian framework, and is presented in an accessible manner in \citep{Ashton:2017ykh}.

\paragraph{\textbf{EM-Bright}} \label{alert}
EM-bright is a pipeline designed to assess whether a \ac{GW} candidate is capable of producing an electromagnetic counterpart \citep{ChGh2020}. A rapid assessment of EM-bright properties, \hasns~and \hasremnant, is essential to trigger \ac{ToO} follow-up by ground and space-based
observatories. In this regard, {\hasns} and {\hasremnant} quantities are reported as a part of the automated and update
discovery notices. The {\hasns} is the probability of the binary having a \ac{NS} component, while {\hasremnant} is
the probability of the merger leaving remnant matter post-merger in the form of dynamical or tidal ejecta.

The exact nature of EM emission from the merger is complex and depends on several factors like the properties of the ejecta,
the \ac{NS} \ac{EoS}, and the \ac{BH} mass and spin. Detailed analyses are required to assess the accurate
properties of EM counterparts (see, for example, \citet{shibata2019} for a review). These are, however, impractical in a real-time
setting. Aside from theoretical uncertainties, measurement uncertainties predominantly affect the assessment of EM-brightness in
real-time. Note that the only real-time data product available from match filter CBC searches is the template parameters that
maximize the likelihood of detection. Bayesian parameter estimation from computationally cheap waveform
models may be available in $\sim$ hours, as discussed later, but it is not available in the seconds after a trigger is registered.
Hence, inference from the template parameters and real-time detection statistics is the feasible solution. 

To this end,
\citet{ChGh2020} showed the application of supervised machine-learning trained on a feature space involving the template
parameters and detection statistics to make this inference. Training is done using large-scale simulation campaigns where the
ground truth and the recovery of search pipelines are registered. The ground truth is labeled based on its intrinsic source-frame
mass as having a \ac{NS} component, or both mass and spin components as leaving remnant matter behind post merger, based on a
phenomenological fit to numerical relativity simulations by~\cite{FoHi2018}. The \ac{NS} \ac{EoS} plays a crucial role in the labeling
as stiffer \ac{EoS} favor tidal disruption, and therefore prefer larger ejecta masses. While in \citet{ChGh2020} a single, stiff \ac{NS} \ac{EoS}
was used on conservative grounds, here we extend the analysis to multiple \ac{EoS}s, and reweight the score based on Bayes
factors computed against GW170817~\citep{AbEA2018c} tidal deformability measurements for several literature \ac{NS} \ac{EoS}s
presented in~\cite{Ghosh2021}. The score presented is therefore marginalized over several \ac{EoS}s.

In addition to {\hasns} and {\hasremnant}, a new quantity {\hasmassgap}, the probability that at least one component of the
binary merger is in the lower mass-gap region i.e. source-frame mass between $3\,\msun$ to $5\,\msun$ is computed. The technique
used in computing {\hasmassgap} is similar to the original EM-bright quantities, except the labeling is different and does not
involve the knowledge of the \ac{NS} \ac{EoS}. The values reported for {\hasns} and {\hasremnant} use a nearest-neighbor classifier algorithm,
while that used for {\hasmassgap} use a random-forest classifier algorithm. The dataset used for training contains additional mass-gap injections done separately on \ac{O2} dataset whereas the feature space used to train the algorithm is the same as~\citet{ChGh2020} -- a five-dimensional space involving the triggered template masses $m_{1,2}$, the
aligned dimensionless spins, ${\chi}_{1,2}$, and the network \ac{SNR}. The duration from which the mass-gap injections were taken from is shown in Table \ref{table:mass-gap}.
\begin{deluxetable}{cc}
\tablehead{
\multicolumn{2}{c}{\textbf{GstLAL Chunks used for Training \hasmassgap~Classifier}} \\
\colhead{Start date} & \colhead{End date}}
\startdata
Sun 2017-01-22 08:00:00 UTC  &  Fri 2017-02-03 16:20:00 UTC \\
Tue 2017-02-28 16:30:00 UTC  &  Fri 2017-03-10 13:35:00 UTC \\
Fri 2017-06-30 02:30:00 UTC  &  Sat 2017-07-15 00:00:00 UTC \\
Sat 2017-08-05 03:00:00 UTC  &  Sun 2017-08-13 02:00:00 UTC \\
Sun 2017-08-13 02:00:00 UTC  &  Mon 2017-08-21 01:05:00 UTC \\
\enddata
\caption{
Calendar times for the detector chunks of LIGO
O2 data. We consider the mass-gap injections performed by the GstLAL search in these duration along with previously existing set in~\citet{ChGh2020} for the study.}
\label{table:mass-gap}
\end{deluxetable}

Similar to the sky maps, these quantities are updated from online parameter estimation samples, which are made publicly available
$\sim$ hours after discovery.
The parameter estimation samples allow for these quantities to be computed directly. 


\subsection{Low-latency Parameter Estimation}\label{PE}

CBC signal candidates labeled as significant (see \ref{significant}) are further investigated via automated Bayesian parameter estimation analysis with the \texttt{Bilby} library. It employs the nested sampling technique implemented in the \texttt{Dynesty} library \citep{Speagle:2019ivv} to explore the full parameter space of masses and spins, producing accurate inference results immune to biases included in the point estimates of masses and spins from search pipelines. It also takes into account uncertainties in detector calibration and marginalizes the posterior probability distribution over them. 

To accelerate the analysis, we employ the reduced order quadrature (ROQ) technique \citep{Canizares:2014fya, Smith:2016qas, Morisaki:2020oqk}, which approximates gravitational waveform with ROQ basis elements to reduce the computational cost of likelihood evaluations.
The ROQ basis elements employed in the automated parameter estimation of O4 are presented in \cite{Morisaki:2023kuq}. 

For \ac{BNS} candidates, the analysis assumes that dimensionless spins have norms less than $0.05$ and are aligned with the orbital angular momentum, and employs the \texttt{IMRPhenomD} waveform approximant \citep{Husa:2015iqa, Khan:2015jqa} to recover the observed signals.
With the acceleration technique, the sampling completes typically in less than 10 minutes. The actual time from detection to upload of results from \texttt{Bilby} is a few tens of minutes since this analysis starts around 5 minutes after signal detection, and preparing input data and post-processing results take several minutes. Currently, the output of this analysis is not automatically made public but manually sent after it passes human vetting. Hence it is sent at the earliest when an Initial Notice is sent, and the actual latency of the update is higher than the latency of upload of the results. 
In addition to this automated analysis, more costly manual analyses incorporating general spin configurations and tidal deformation of colliding objects may follow, depending on the significance of the signal.
For candidates with higher masses, the automated analysis takes into account general spin configurations, and employs \texttt{IMRPhenomXPHM} \citep{Pratten:2020ceb} if its ROQ basis elements are available in the target mass range, and \texttt{IMRPhenomPv2} \citep{Hannam:2013oca} otherwise. 
This analysis typically takes hours to complete. The output of this analysis is released as an Update Notice. The UV-optical radiation from a kilonova is expected to fade away within $\sim$ 48 hours \cite{AbEA2017f}, so parameter estimation updates within $\sim$ hour are sufficient for follow-up purposes.


\section{Mock Data Challenge}
\label{sec:mdc}

\begin{deluxetable*}{cccccc}
\tabletypesize{\footnotesize}
\tablecolumns{8}
\tablewidth{0pt}
\tablehead{
\multicolumn{6}{c}{\textbf{Compact Object Properties}} \\ [-0.1cm]
\colhead{Binary Type} & \colhead{Object} & $m/\msun$ (min/max) & $m$ distribution & Max $a$ & $a$ distribution}
\startdata
BNS & Primary & 1.0 - 2.05 & uniform & 0.4 & uniform \& isotropic \\ [-0.1cm]
&  Secondary & 1.0 - 2.05  & uniform & 0.4   & uniform \& isotropic \\[0.1cm]
NSBH & Primary & 1.0 - 60.0 & $m^{-1}$ & 0.998 & uniform \& isotropic \\ [-0.1cm]
& Secondary & 1.0 - 2.05  & uniform & 0.4   & uniform \& isotropic \\[0.1cm]
BBH & Primary & 2.05 - 100 & $m^{-2.35}$ & 0.998 & uniform \& isotropic \\ [-0.1cm]
&  Secondary & 2.05 - 100  & $m^1$ & 0.998   & uniform \& isotropic \\
\enddata
\caption{Distribution of intrinsic properties (component masses $m$ and spins $a$) of binary systems in the injection sample. The spin distributions are uniform in magnitude and isotropic in orientation, as seen in the last column.
\label{table:inj_dist}}
\end{deluxetable*}




To create a background for the \ac{MDC}, we consider the stretch of data taken between Jan 05, 2020 -- Fri Feb 14, 2020 by the \ac{LVK} instruments during O3.  A total of {\totalinjections} simulated CBC
waveforms with mass and spin distributions mentioned in Table~\ref{table:inj_dist} are injected into the O3 data with an interval of $\sim 1$ minute between injections. 
The optimal network \ac{SNR} is greater than 4 for all the injections. This is done
to prevent ``hopeless'' injections, which are improbable to be detected in reality. The {\tt{IMRPhenomPv2\_NRTidalv2}}
waveform approximant is used to consider matter effects in case of \ac{NS} components of the
injections. For \ac{BH}s, the same waveform is used with the tidal parameters set to zero. In order to label a component as a \ac{NS}, the SLy~\citep{ChBo1998} \ac{NS} \ac{EoS} is used, which allows
for a maximum mass of $\sim 2.05\;\msun$. Hence, in this scheme, the tidal deformability of component masses above this limit
are set to zero consistent with being a \ac{BH}. In particular, all injections above the SLy maximum mass are assumed to be
\ac{BH}s, and the appropriate relative rate is used for the same.
These injected signals are primarily recovered by CBC pipelines, and occasionally by Burst pipelines. In the \ac{MDC} exercise, 
we have focused most of our analysis on the output and data products of the CBC pipelines.
We also note that the injection rate density used in the study
is artificially high and not representative of the true discovery rate in O4. We expect $\mathcal{O}(10^2)$ \ac{CBC} detections during the full duration of \ac{O4} \cite{Kiendrebeogo:2023hzf}, compared to $\mathcal{O}(10^3)$ of detections across the 40 day \ac{MDC} cycle. Therefore, quantities like \pastro which rely
on the background distribution, may not be the true representation as compared to a realistic signal density. 
The CBC injection set consists of {\BNSinj} \ac{BNS}, {\NSBHinj} \ac{NSBH}, and {\BBHinj} \ac{BBH} injections.



The events are distributed uniformly in co-moving volume assuming flat $\Lambda$CDM cosmology with $H_0 = 67.3\;\hubbleunit$ and $\Omega_m = 0.3$ based on Planck 2018 results mentioned in Table 1 of \cite{planck2018}. 
The \ac{BNS} systems are distributed up to a maximum redshift of $z = 0.15$, the neutron-star black-hole
systems up to $z = 0.25$, and the \ac{BBH}s up to $z = 1.9$. The simulated strain is projected on to the detector geometries,
shifted in time to the time of experiment and streamed as 1\,s segments for the search and annotation pipelines to analyze in
real-time (see Section~\ref{sec:searches}). 
The triggers and their annotations were reported in
\ac{GraceDB} for post processing studies.

This exercise is repeated in several cycles for benchmarking analysis and will continue internally during the observing run to continuously track improvements in the alert infrastructure and provide avenues for pipelines to test their changes. The numbers reported here are those from a single cycle of \totalinjections ~injections where the status of most analyses were close to their \ac{O4} configurations.
 
\section{Results} 
\label{sec:results}

\subsection{Pipeline Performance}

In order to demonstrate LLAI readiness for O4, we compare \ac{CBC} triggers and their corresponding data products uploaded to \ac{GraceDB} with the injection set. 
Triggers and injections are matched using the merger time; all triggers within 1\,s of an injection are matched to that injection and included in the analysis. 
This process removes noise triggers between injections and ensures trigger times correspond to an injection time window. It is possible for a few noise triggers to be coincident with injection time window. Matching triggers to injections allows us to evaluate data products by comparing the results with the injected quantities. 
These data products include basic parameter estimates, sky maps, and \pastro values. However, not all injections are recovered in the form of trigger. 
There are four main reasons why an injection may not be found by the search pipelines: (i) some injections are distant or have a low \ac{SNR}, due to the cosmological distribution preferring larger distances, and can not be distinguished from background noise, (ii) there are stretches of the O3 replay where one or more detectors were not operational and in science mode, (iii) there were some temporary technical issues on computing resources used during this MDC cycle, and (iv) there may be data quality issues that overlap with an injection, such as loud or long glitches. 




\begin{figure}[ht!]
\centering
\includegraphics[scale=.40]{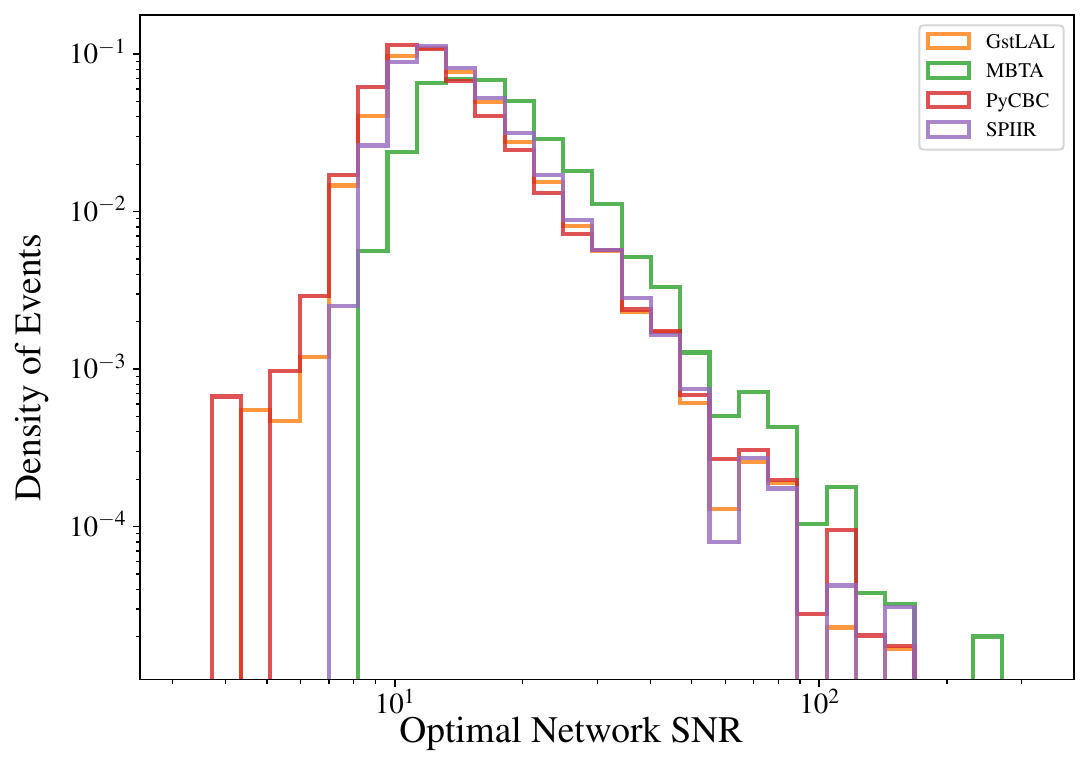}
\caption{Histograms of the  optimal, or injected, network \ac{SNR}, normalized for each CBC pipeline, for triggers below the significant FAR threshold.  All pipelines were found to recover injections across the range of injected \ac{SNR} values.
\label{fig:snr_hist}}
\end{figure}

\begin{figure}[ht!]
\includegraphics[scale=0.5]{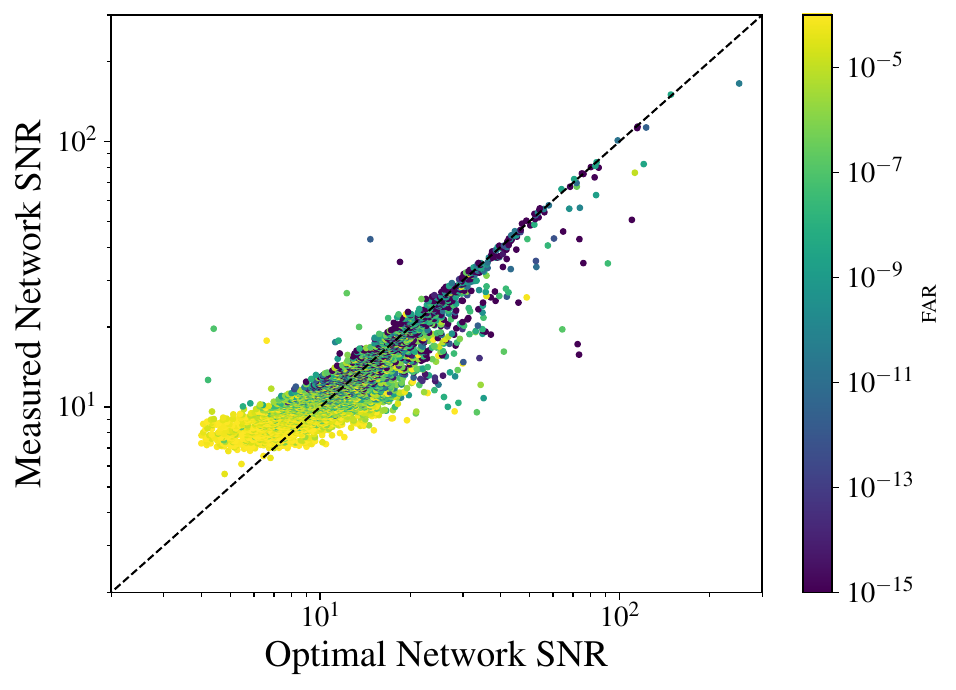}
\caption{The measured network \ac{SNR} recovered during the MDC compared to the optimal, or injected, network \ac{SNR} with the points colored by FAR. We find \ac{SNR} is recovered more accurately for higher values.}
\label{fig:SNR_inj_rec}
\end{figure}

As mentioned in the previous section, for this analysis we focus on the MDC cycle used for the review of pipeline performance, which ran from February 16 through March 28, 2023 consisting of \totalinjections \ injections. 
During this MDC, {\BNSrec} \ac{BNS}, {\NSBHrec} \ac{NSBH}, and {\BBHrec} \ac{BBH} injections were recovered.
As seen in Figure~\ref{fig:snr_hist}, each of the \ac{CBC} search pipelines, \pycbc, \ac{GstLAL}, \ac{MBTA}, and \ac{SPIIR}, successfully uploaded events below the significant FAR threshold across the range of injected \ac{SNR} values. Burst searches, \ac{cWB} and oLIB, make little assumptions of source type, and so are only considered for measures of latency in this paper.
We plot the simulated vs.\ recovered network \ac{SNR} in Figure~\ref{fig:SNR_inj_rec}. In general, signals with moderate to high \ac{SNR} are recovered well, with some bias at low \ac{SNR}s due to the FAR threshold imposed for upload. 
Additional scatter in \ac{SNR} recovery is expected since the simulated optimal \ac{SNR}s were calculated using fixed detector sensitivities, whereas actual detector data has significant fluctuations in sensitivity over time. The optimal \ac{SNR}s for injections are calculated using a global average PSD, instead of using a local estimate of the PSD, which may cause some of the off-diagonal outliers.

For the 4514 GW injections found, we created 469 
multimessenger coincidences by injecting simulated GRB candidates at times surrounding the GW injections.  We found 356 of these joint candidates triggered a RAVEN alert as a result of passing the \textit{significant} FAR threshold.

\subsection{Latency Measures}


Due to the desire for timely follow-up by the multi-messenger community, a key feature of the LLAI is dissemination of results as quickly as possible.
The goal for the LLAI system is to send alerts for events within 30\,s of merger time; this number sets the timescale for comparison below.
Here, we perform a systematic study of the alert latency for three of the key pieces of the pipeline (a fourth, the data calibration, construction, and transfer between sites, which takes $\sim$ 5-10\,s, is not captured here, as well as latency from the ingestion and redistribution by \acs{GCN} or \acs{SCiMMA}).
Latency comes primarily from these three components: (i) the search pipelines, (ii) the event orchestrator \gwcelery, and (iii) \ac{GraceDB}.
We note that technical issues during the MDC may also cause some high-latency outliers, so the results presented are conservative when excluding the time needed for transfer and construction of the strain data.

\begin{deluxetable}{cCCC}
\tablehead{
\colhead{Latency Measure} & \colhead{Description} & \colhead{50\% (s)} & \colhead{90\% (s)}
}

\startdata
Superevents & \tsuperevent - \tmerger & 9.4 & 18.1 \\
CBC Events & \tevent - \tmerger & 12.3 & 41.4 \\
Burst Events & \tevent - \tmerger & 72.3 & 671.3 \\
Early Warning Events & \tevent - \tmerger & -3.1 & 2.9 \\
GW Advocate Request & \tadvreq - \tmerger & 12.7 & 40.1 \\
GCN Preliminary Sent & \tgcn - \tmerger & 29.5 & 171.8 \\
Coincidence with GRB Found & \temcoinc - \tmerger & 32.9 & 44.4 \\
RAVEN Alert Triggered & \travenalert - \tmerger & 35.3 & 48.4 \\
\enddata
\caption{Measured latencies for a number of steps in the pipeline. $t_0$ corresponds to the event merger time reported by the pipeline, while $t_{superevent}$ and $t_{event}$ correspond to the time of superevent or event creation. For the case of superevent latencies, $t_0$ is determined by the preferred event.}
\label{table:latencies}
\end{deluxetable}

\begin{figure*}[ht!]
\includegraphics[scale=0.5]{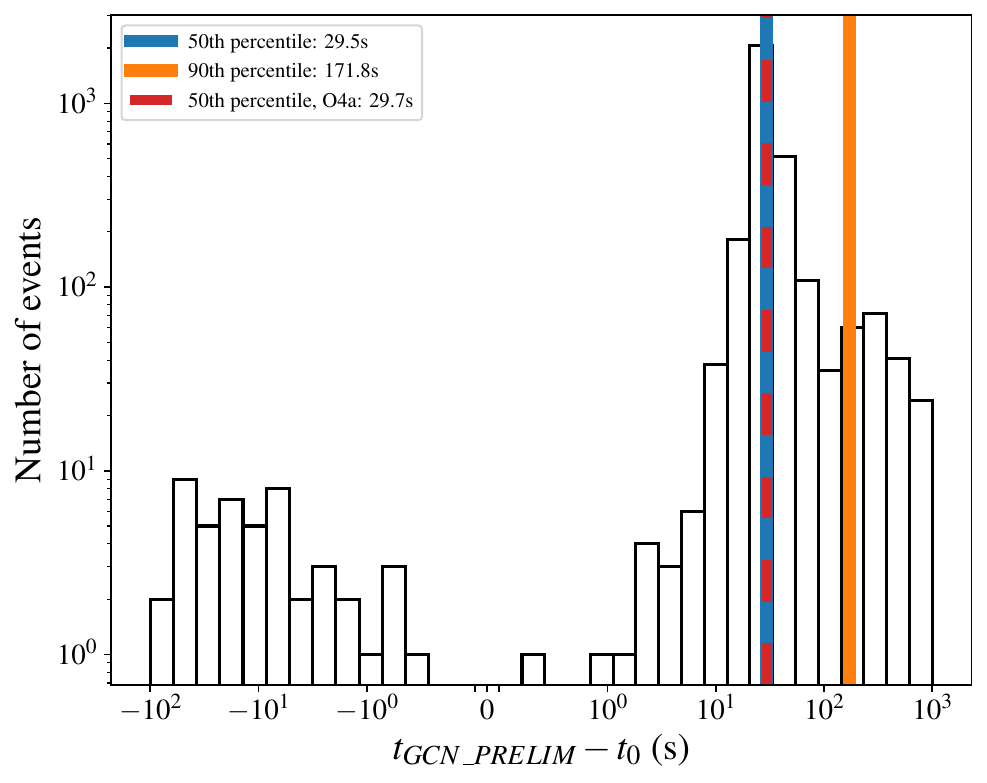}
\includegraphics[scale=0.5]{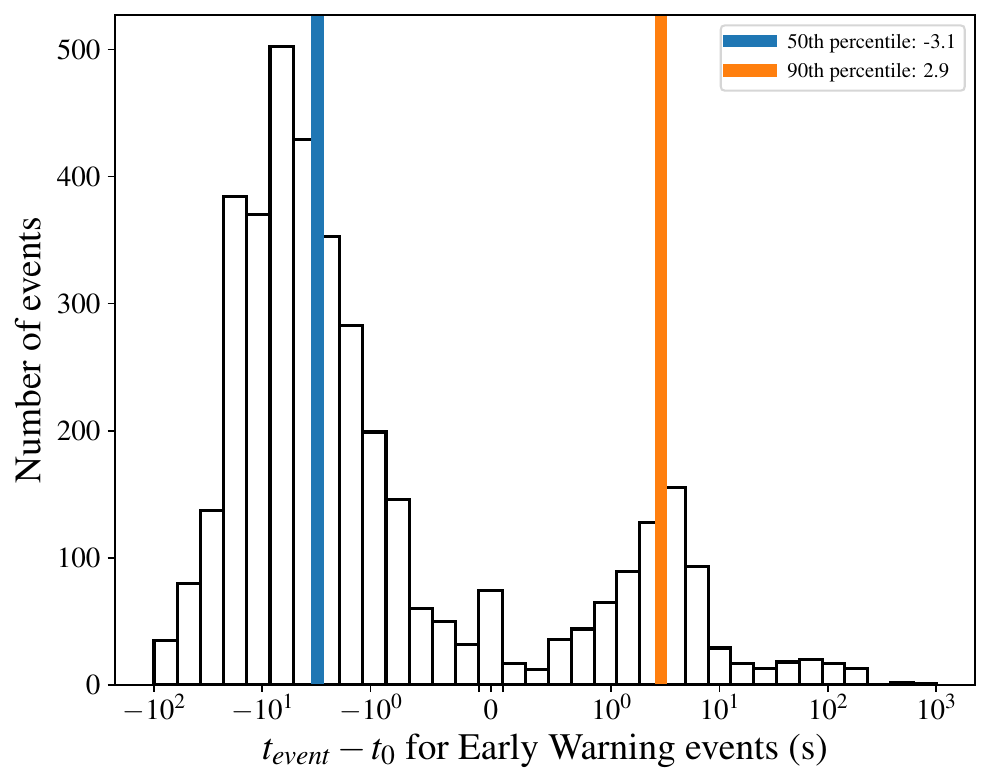}
\caption{\textit{Left}: Histogram of latencies for the sending of the GCN preliminary alert. We compare the \ac{MDC} latencies to the median latency measured during \ac{O4}a. The \ac{O4}a measurement includes data calibration, construction, and transfer time, while the \ac{MDC} latencies do not. \textit{Right} Histogram of latencies for Early Warning alerts. $t_{GCN\_PRELIM}$ corresponds to the time the GCN preliminary alert is sent, $t_0$ corresponds to the preferred event merger time, and $t_{event}$ corresponds to the time of event creation.}
\label{fig:latencies}
\end{figure*}

To calculate event latencies, we compare the time an event is created and appears on \ac{GraceDB} to that of the known merger time. For \ac{GW} event candidates during an observing run, the merger time is defined as the time the signal peak reaches Earth's center.
For CBC pipelines, we find a median (90\%) latency of 12.3\,s (41.4\,s); for Burst pipelines, we find a median (90\%) latency of 72.3\,s (671.3\,s)
We then compare this number to the creation of a superevent; we find a median (90\%) latency of 9.4\,s (18.1\,s). The median superevent latency is lower than the event latency simply due to the fact that the superevent may be created upon the first trigger, and that a superevent often consists of multiple events.
We also make the same measurement for Early Warning alerts, shown on the right of Figure~\ref{fig:latencies}; we find a median (90\%) latency of -3.1\,s (2.9\,s). Considering the joint candidates, we find that it takes a median (90\%) latency of 32.9 \,s (44.4\,s) to find a coincidence with a GRB injection and a median (90\%) latency of 35.3 \,s (48.4\,s) to trigger a RAVEN alert.

Once the event(s) have been created, there is a request for human vetting of the alert, called the Advocate Request; we find a median (90\%) latency of 12.7\,s (40.1\,s) to notify the advocate.
To measure the latency of event communication to the community, we also measure the latency for sending of the GCN preliminary alert, which occurs for superevents that pass automated data quality checks; we find a median (90\%) latency of 29.5\,s (171.8\,s). We show this statistic for GCN preliminary alerts on the left of Figure~\ref{fig:latencies}. This latency reported specifically measures that time until the GCN preliminary label is applied. We also compare this to the measured median GCN latency during \ac{O4}a, the first half of \ac{O4}, which, including data calibration, construction, and transfer time, is 29.7\,s. The agreement between the GCN latency during the \ac{MDC} and \ac{O4}a shows our latency has not increased, and has likely slightly decreased as the \ac{O4}a measure includes data calibration, construction, and transfer time, while the \ac{MDC} measurement does not.
Table \ref{table:latencies} shows a compilation of these latency statistics for comparison.
The number of candidate events within a superevent was not shown to have a noticeable effect on the latency of that event and its corresponding preliminary alert.

\subsection{Probability of astrophysical origin}
\label{subsec:pastro_results}

\begin{figure}[ht!]
\centering
\includegraphics[width=\columnwidth]{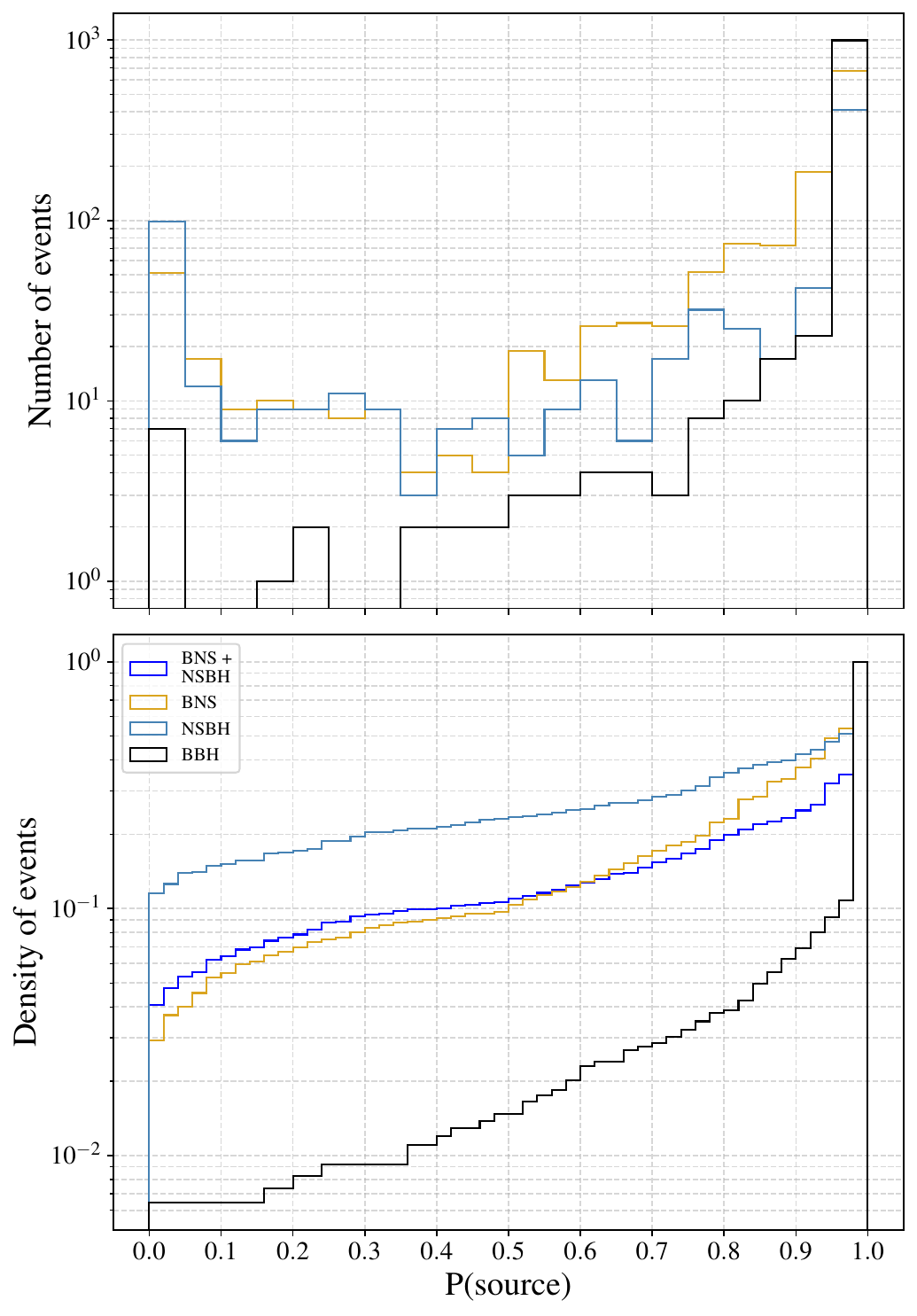}
\caption{\textit{Top:} Histogram of ``preferred'' recovered P(source) for true sources, for superevents which pass the significant public alert threshold.
The possible source classes are \ac{BNS}, \ac{NSBH}, and \ac{BBH}.
This excludes early warning events, which were not fully functional during the time of this analysis.
\textit{Bottom:} Cumulative density of the same data. We also include the distribution of correctly recovered \ac{BNS} or \ac{NSBH}, which checks for contamination between the two classes due to misrecovered secondary masses or effects from varying the \ac{NS}/\ac{BH} mass boundary. As this is a cumulative histogram, the fraction of events above a certain P(source) corresponds to the \ac{TPR}. We see that the majority of events with a P(source) greater than 0.5 correctly recover the injection source type. 
\label{fig:psource}}
\end{figure}

Each CBC pipeline uploads its own estimate of \pastro, as described in Section~\ref{sec:searches}, and is inherited by the superevent if the event is the preferred event. By matching the MDC's injected parameters to the recovered \pastro values from pipelines, we can test the accuracy of \pastro.
Figure~\ref{fig:psource} shows the recovered P(source) for true sources (e.g., P(BNS) for true injected \ac{BNS} events), for superevents which pass the public alert FAR threshold.
In matching these injections, we place the cut between \ac{NS} and \ac{BH} at 3 $M_\odot$.
``True'' terrestrial events correspond to superevents which were not temporally matched to injections, indicating that they arose from detector noise.
Injected \ac{BBH}s are typically recovered confidently, with the vast majority resulting in P(BBH) $>$ 0.5. Over 90\% of \ac{BBH}s are recovered with P(BBH) $>$ 0.9.
The P(BNS) and P(NSBH) distributions are less confident than the P(BBH) distribution; about 10\% of true \ac{BNS}s and \ac{NSBH}s are recovered with P(BNS) or P(NSBH) $< 0.1$.
As a check against contamination across P(BNS) and P(NSBH) due to errors in recovered masses or variation in the definitions of the mass border between \ac{NS} and \ac{BH} (i.e., the Tolman–Oppenheimer–Volkoff mass), we can also look at the distribution of the sum of P(BNS) and P(NSBH) for injected events where $m_2 \leq 3 M_\odot$. 
This is shown in the bottom panel of Figure~\ref{fig:psource}. Compared to P(BNS) or P(NSBH) alone, P(BNS) + P(NSBH) performs slightly better, with $\sim 75\%$ of true \ac{BNS} or \ac{NSBH}s receiving P(BNS) + P(NSBH) $> 0.9$, compared to 60\% and 65\% for \ac{NSBH} and \ac{BNS} respectively. 
If instead we use a threshold P(source) of $.5$, we find a \ac{TPR} of $\sim 98$ \% for \ac{BBH}, $\sim 90$ \% for \ac{BNS}, and $\sim 76$ \% for \ac{NSBH} injections.

\subsection{Sky localization} \label{subsec:skymapresults}

In order to evaluate localization performance, in the following, we focus on three metrics: (i) localization area, (ii) retrieved median distance of the source, and (iii) searched area. Localization area is the area (measured in$~\mathrm{deg^2}$) which encloses a given probability contour in the sky map; in this paper, we use 90\% as the total cumulative probability threshold. Retrieved distance refers to the median of the distance distribution along the line of sight of the injected sky position. Searched area is the smallest 2D area, starting with the regions of highest probability, that contains the true location of the source; it represents a measurement of the sky area that a telescope with a small FOV relative to the sky map size would need to cover before imaging the true location. We refer the reader to \cite{2016PhRvD..93b4013S} for more details on these parameters, and use the \texttt{ligo.skymap}\footnote{\url{https://git.ligo.org/leo-singer/ligo.skymap}} package to compute all metrics. 

\begin{figure}[ht!]
\centering
\includegraphics[width=\linewidth]{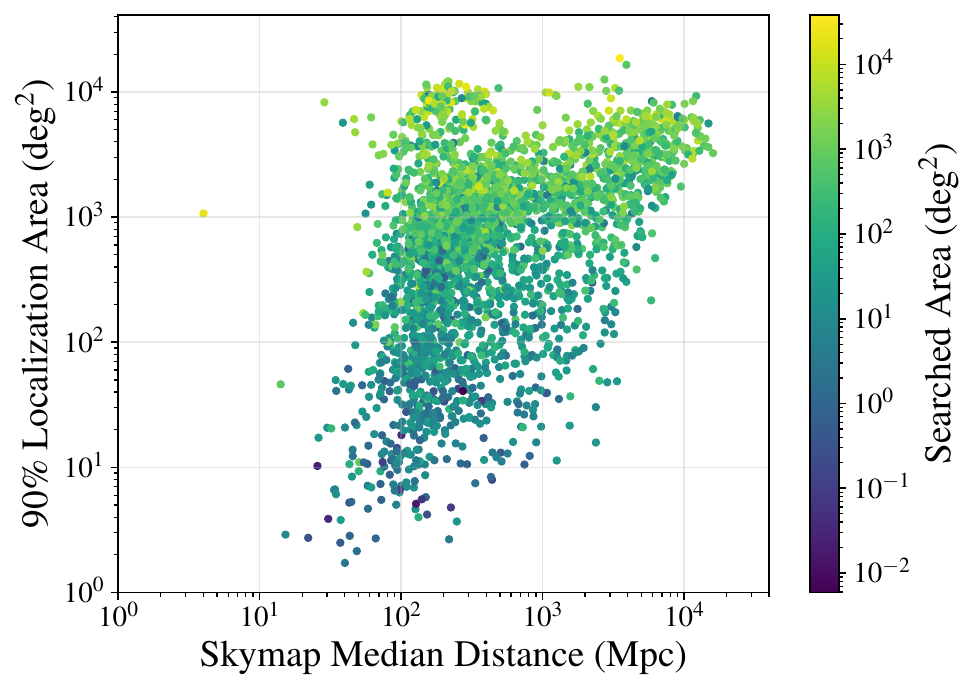}
\caption{Sky localization distribution as a function of recovered median distance from the sky map, with the trend of Searched Area for the preferred event. We find the more distant the event, the larger is  the localization area. The color bar shows that the searched area associated with the event also increases with the localization area and sky map median distance as discussed in Section \ref{subsec:skymapresults}.}
\label{fig:skymap_loc}
\end{figure}

\begin{figure*}[ht!]
\centering
\includegraphics[scale=.45]{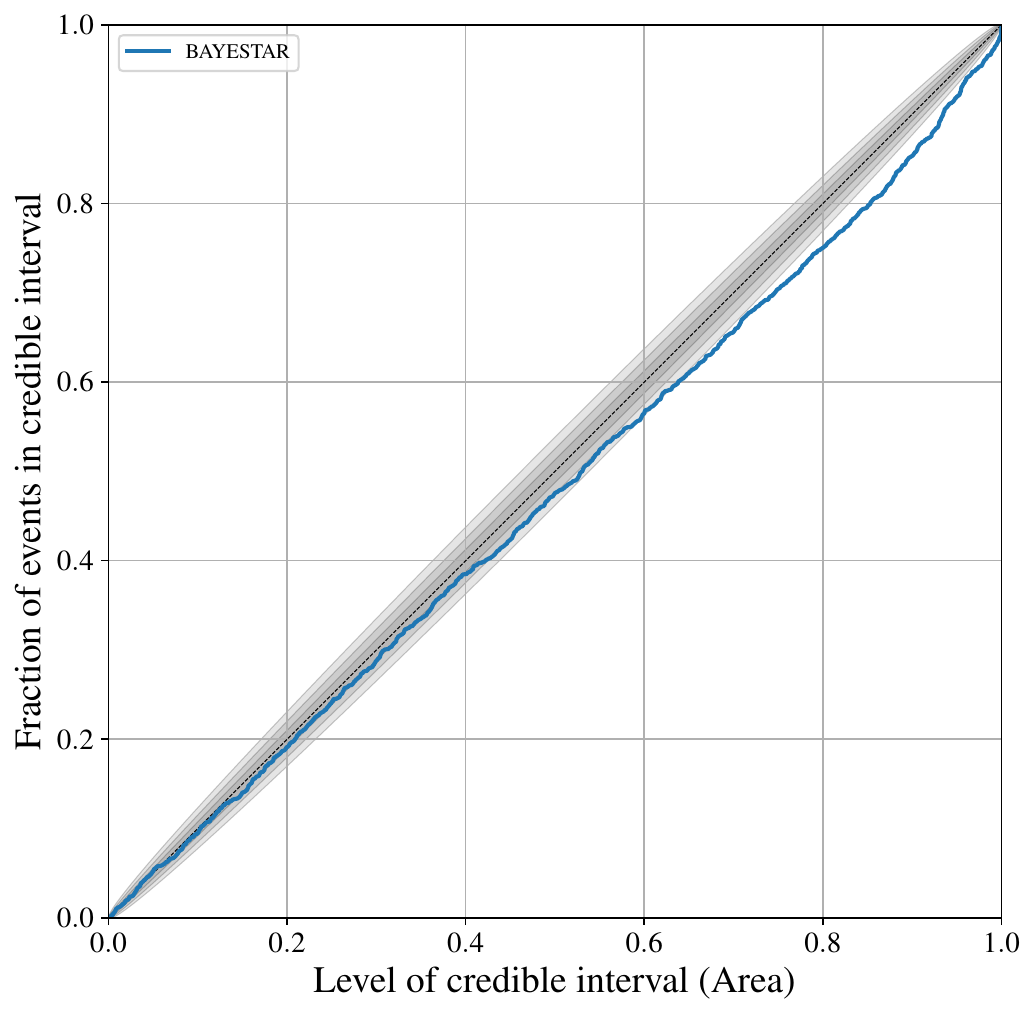}
\centering
\includegraphics[scale=.45]{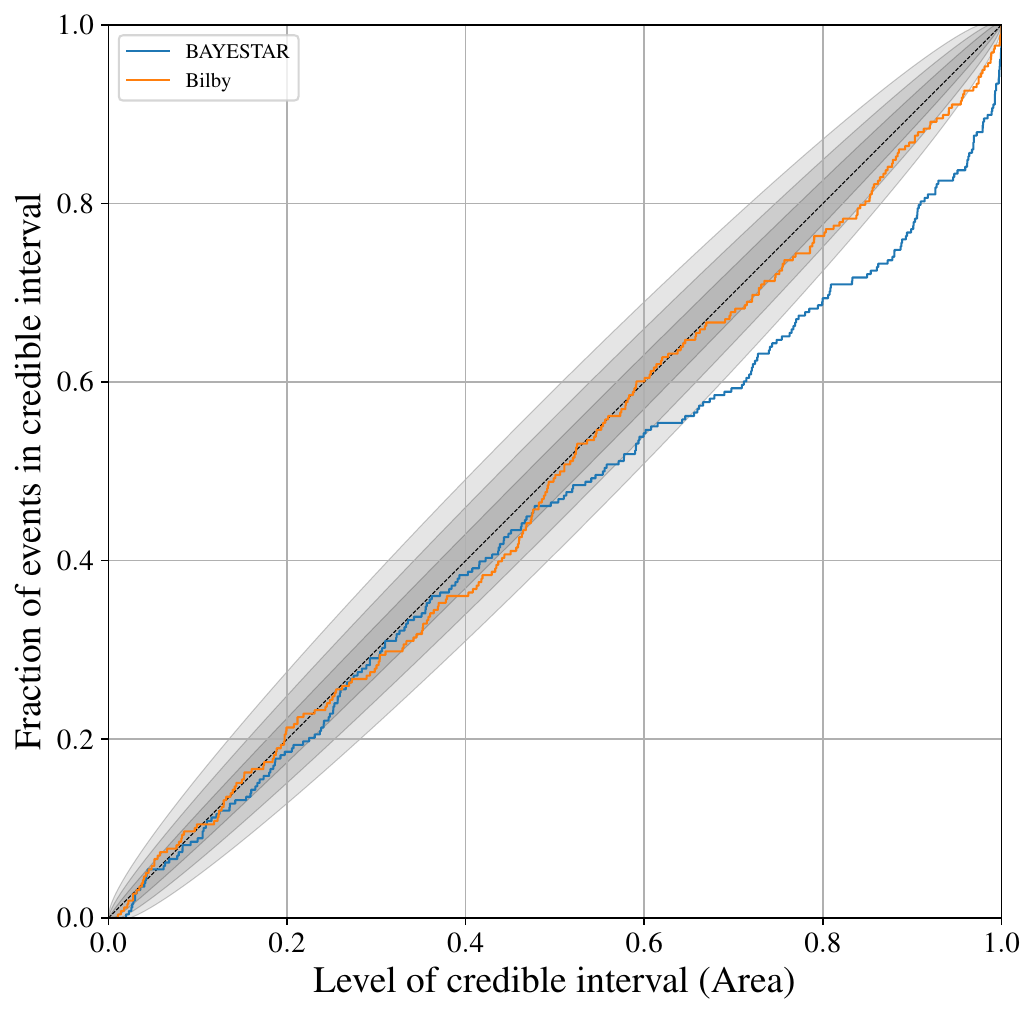}
\centering
\includegraphics[scale=.45]{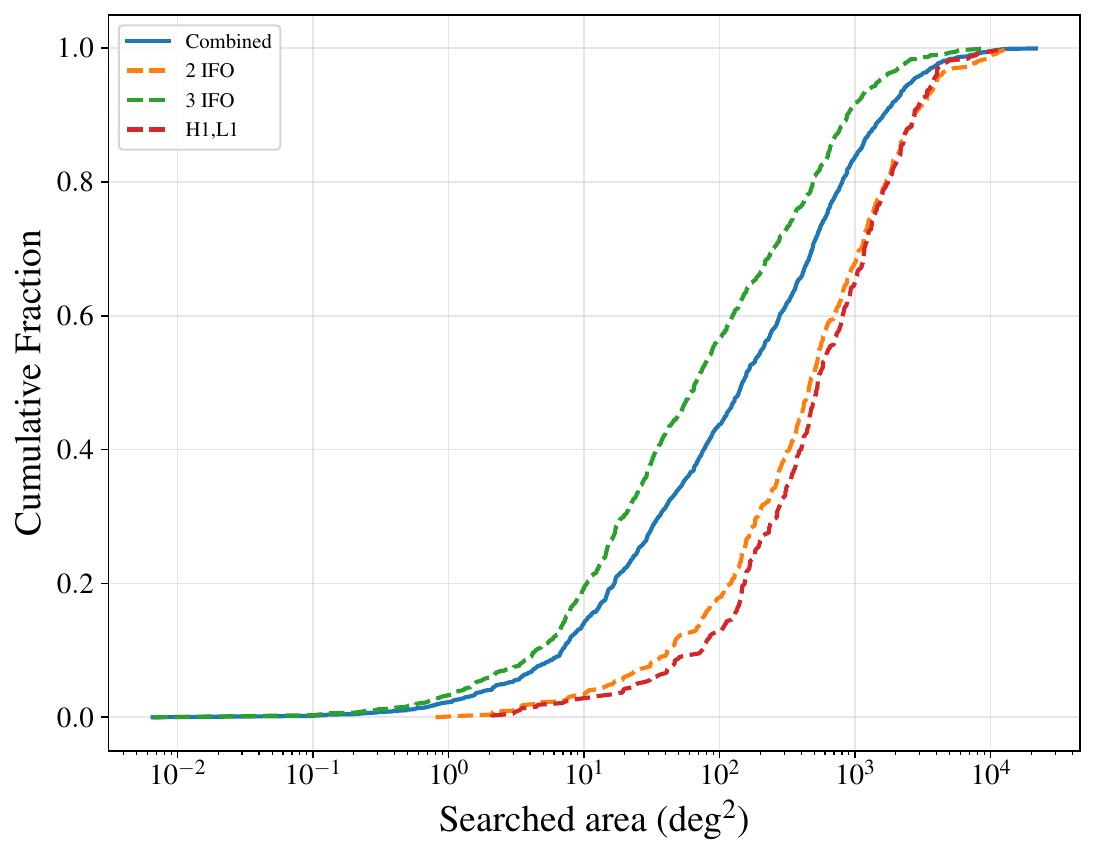}
\centering
\includegraphics[scale=.45]{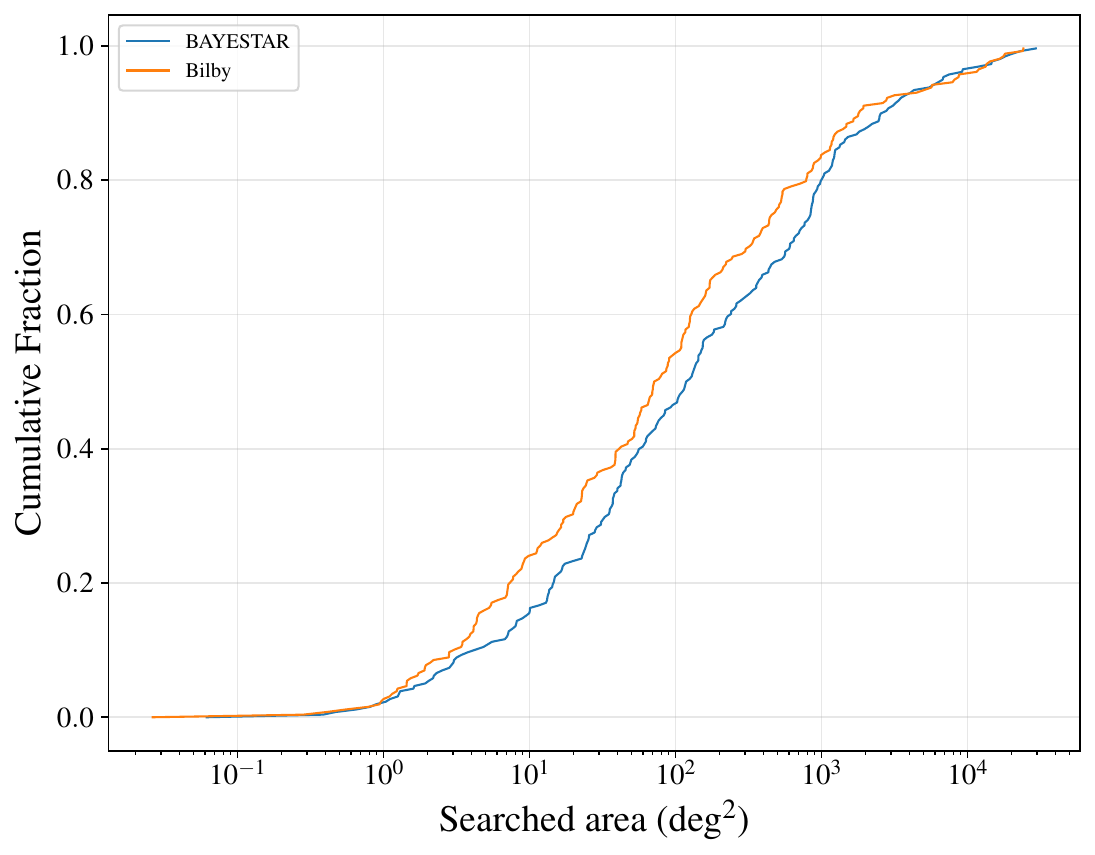}
\caption{\textit{Top left:} A \textit{P-P} plot showing the \texttt{BAYESTAR} sky map statistics for the preferred event. The credible intervals shown in gray are based on the total number of events. \textit{Bottom left:} Cumulative histograms of \texttt{BAYESTAR} searched area for all events (blue), compared to two-interferometer (Orange), three-interferometer (Green), and Hanford (H1) and Livingston (L1) (Red) events. We see that the three-interferometer events produced smaller searched areas than the two-interferometer events by almost an order of magnitude as discussed in Section \ref{subsec:skymapresults}. 
\textit{Top right:} A \textit{P-P} plot showing the performance of \texttt{BAYESTAR} (blue) and \texttt{Bilby} (orange) generated sky maps for \ac{BNS} events. The credible intervals shown in gray based on the total number of such preferred events where both \texttt{BAYESTAR} and \texttt{Bilby} sky maps are available. \textit{Bottom right:} Cumulative histograms showing searched area statistics for \texttt{BAYESTAR}  and \texttt{Bilby} sky maps. We observe that \texttt{Bilby} sky maps give a lower searched area and tend to be more precise than their \texttt{BAYESTAR} counterparts as discussed in \ref{subsec:skymapresults}.}
\label{fig:skymaps}
\end{figure*}

When evaluating sky map performance, we consider the preferred events for superevents that fall under the significant FAR threshold before trials factor and were detected by more than one interferometer.
We exclude single detector triggers as most resulted from injections that occurred during a portion of O3 replay data where one or more detectors was not in science mode, and they may have sky localizations on the order of the entire sky.
For Figure \ref{fig:skymaps}, We also exclude a slice of parameter space for injected NSs of mass $\leq 2 \msun$ with spins $\geq .05$, as these events may not have a match within the pipeline template banks.
We compare the 90\% localization area with the recovered distance in Figure~\ref{fig:skymap_loc}. We find a positive correlation between the retrieved distances of the injection and the localization areas, as in general greater distances lead to larger localizations. This same trend was seen for the searched areas for these superevents, where at greater distances one would typically encounter larger searched areas, which aligns with the expected behavior.


Further, in the top left panel of Figure~\ref{fig:skymaps}, we show the accuracy of \texttt{BAYESTAR} sky maps through a \textit{P-P} plot. \textit{P-P} plots of this format show the fraction of injections found within a given credible interval across all levels of credible intervals. The three gray lozenges around the diagonal shows the three different levels of confidence (1-3 $\sigma$) for the combined \texttt{BAYESTAR} map sample. We find that the \texttt{BAYESTAR} sky maps fall just outside the credible intervals for higher credible intervals. This tells us \texttt{BAYESTAR} slightly overstates the precision of its sky localizations. We also show the performance of different pipelines that detected the preferred event in the given sample. See the appendix for a discussion of sky map performance for each individual pipeline. In the bottom left panel of Figure~\ref{fig:skymaps}, we show the cumulative trend of the searched area statistics from the combined sample. We find a median searched area of $~100-200~\mathrm{deg^2}$. We then compare the two interferometer events with the three interferometer events, to show that the latter produced smaller searched areas. We also include a line for the two-interferometer case where Hanford (H1) and Livingston (L1) specifically observed the event, as those are the interferometers currently being used during \ac{O4}. The \acs{LVK} Alert User Guide\footnote{\url{https://emfollow.docs.ligo.org/userguide/}} provides up-to-date information about the interferometers currently in use. In this case, we see that there is marginal difference between the two-interferometer line, and the H1, L1 line. 

Further, we probed the accuracy of the \texttt{BAYESTAR} sky maps compared with \texttt{Bilby} sky maps for \ac{BNS} events. This comparison \textit{P-P} plot and searched area histogram can be seen in right side of Figure~\ref{fig:skymaps} for the preferred event of all \ac{BNS} superevents for which both sky maps were produced. In the right half of this figure, we include the events excluded in the upper left panel to demonstrate \texttt{Bilby}'s performance even without the cuts. A plot with those cuts applied can be found in the appendix. From the top right panel of Figure~\ref{fig:skymaps}, we see that the \texttt{BAYESTAR} sky maps tend to sag below \texttt{Bilby}'s implying that their precision was overstated as compared to \texttt{Bilby}. There is a trade-off between the latency of \texttt{BAYESTAR} sky maps available with the Preliminary GCN alert, and the improved accuracy of the \texttt{Bilby} sky maps that are available with the completion of parameter estimation. The cumulative searched area plot in the bottom right panel of Figure~\ref{fig:skymaps} shows that typically \texttt{Bilby} sky maps have lower searched area by factor of 2 or more. 

{\color{blue}}

\subsection{EM-Bright}

\begin{figure}[t]
\centering
\includegraphics[width=1.0\columnwidth]{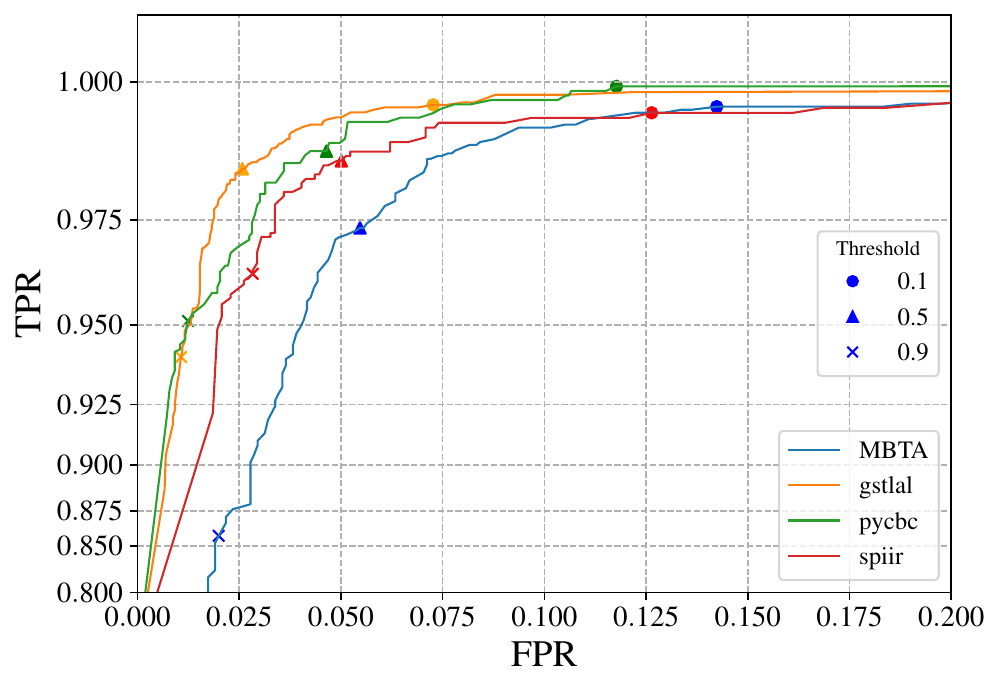}
\\
\includegraphics[width=1.0\columnwidth]{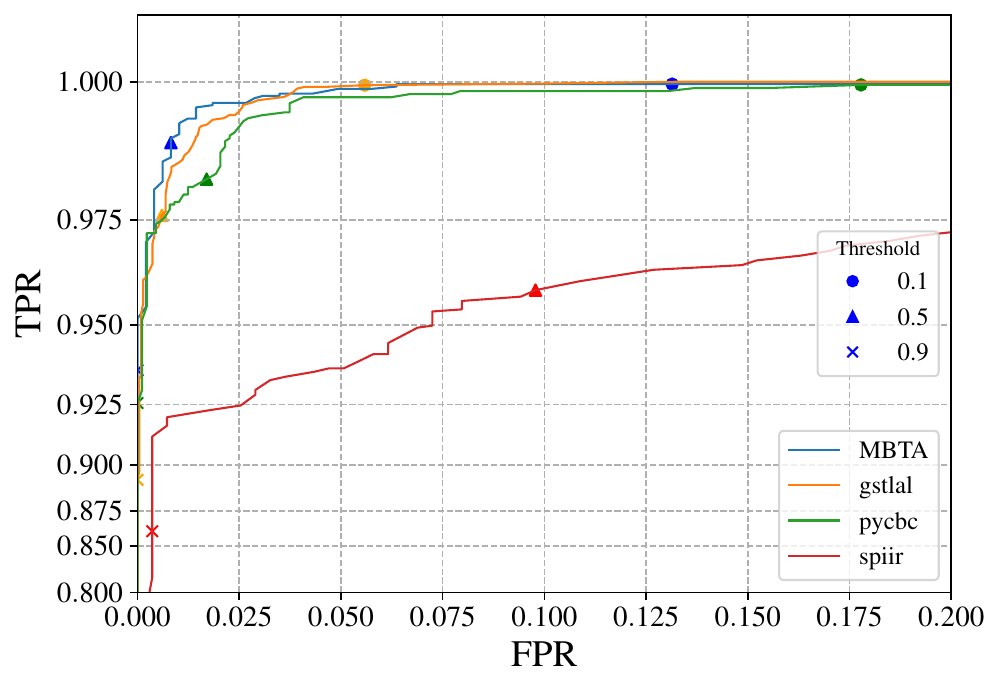}
\\
\includegraphics[width=1.0\columnwidth]{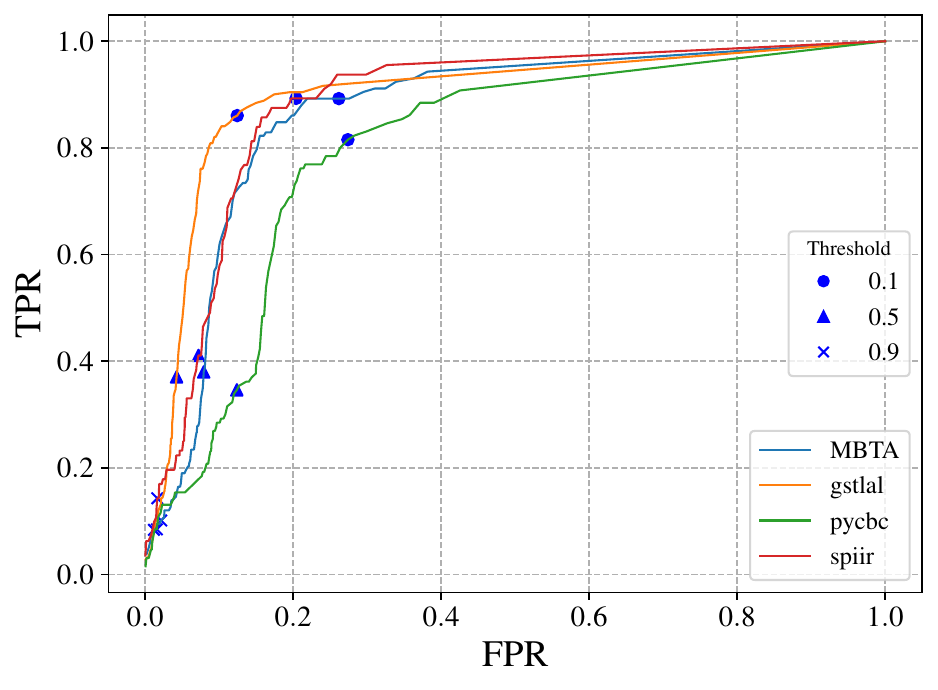}
\caption{The ROC curves for the different EM-Bright
classifiers are shown here for MDC 11 events. The top, middle, and bottom panels refer to {\hasremnant}, {\hasns}, and {\hasmassgap}
quantities respectively. The markers denote different representative thresholds
along the curve.}
\label{fig:sourceproperties}
\end{figure}
We show the performance of the EM-Bright classifiers \citep{ChGh2020} across all four CBC pipelines via their Receiver Operating Characteristic (ROC) curves. For this, the \ac{NSBH} boundary is chosen according to SLy \ac{EoS} but the probabilities are \ac{EoS} marginalized, meaning we include some uncertainty in the \ac{NS} \ac{EoS} in our classifications. The markers indicates three different representative thresholds (a score above which events are considered to be positively classified as the source type in question) along the ROC curves for each pipeline. 
In Figure~\ref{fig:sourceproperties},  we see that the {\hasremnant} quantity for all four pipelines has greater than a 95\% \ac{TPR} for a 5\% \ac{FPR}. In the middle panel, we see (\ac{GstLAL}, \pycbc, \ac{MBTA}) perform consistently at $\sim 97\%$ \ac{TPR} at $\sim 3\%$ \ac{FPR} for {\hasns} classifier. The \ac{SPIIR} pipeline purity is slightly lower compared to the other pipelines, $\sim 94\%$ \ac{TPR} at the same misclassification fraction. One possible mitigation technique is to use more training data from the pipeline in the training process. In the last panel, for {\hasmassgap},  we see (\ac{GstLAL}, \ac{SPIIR}, \ac{MBTA}) perform with $\sim 80\%$ \ac{TPR} at $\sim 20\%$ \ac{FPR} while \pycbc lags slightly below.  We expect to enhance the performance of these classifiers further in the near future by retraining the classifiers using O3 MDC data considering all pipelines.  

\section{Conclusion}
\label{sec:conclusion}

In this paper, we present the performance of the low-latency alert infrastructure and associated data products based on the O3 \ac{MDC}. 
A large simulation campaign of compact binaries i.e. \ac{BNS}, \ac{NSBH}, and \ac{BBH} are injected into a stretch of real data from \ac{O3}. 
The data is taken through the entire end-to-end alert infrastructure starting from the search pipeline, data products computation, and alert generation.
We demonstrate that for full bandwidth searches automated preliminary alerts, excluding time for data transfer and construction, are delivered with a median latency of $\lesssim$ 30\,s, which is an improvement since O3 \citep{LIGOScientific:2021djp}. 
We show that low-mass \ac{BNS} injections are successfully detected by early warning searches. Annotations and alert delivery is achieved for a significant fraction of such signals before merger time with a median alert latency of $\sim$ -3\,s. It is to be noted however, that alert delivery before merger time does not happen for all early warning events. 
In addition, through the use of this MDC dataset, we demonstrate that the data products, produced in the same workflow as planned for O4, are statistically consistent with simulated values.

The \pastro values giving probability of an astrophysical \ac{BBH}, \ac{BNS}, or \ac{NSBH} were found to correctly classify the source for the majority of events. For a threshold P(source) of $0.5$, we find a \ac{TPR} of $\sim 98$ \% for \ac{BBH}, $\sim 90$ \% for \ac{BNS}, and $\sim 76$ \% for \ac{NSBH} injections. 

The distribution of injected sky positions is found to be well recovered by the sky maps produced by both \texttt{BAYESTAR} and \texttt{Bilby}, as evidenced by Figs.~\ref{subsec:skymapresults}. \texttt{BAYESTAR} provides low-latency sky maps that slightly overstate the precision, while \texttt{Bilby} provides improved accuracy upon completion of parameter estimation. The median searched area is found to be $~100-200~\mathrm{deg^2}$, with slight variations between method and pipelines. We observed that \texttt{Bilby} sky maps have better precision, and typically gave a smaller searched area. 

The EM-Bright values corresponding to the probabilities of {\hasns} and {\hasremnant} have a \ac{TPR} of above $\sim 95 \%$ at $\sim 5 \%$ \ac{FPR} across \ac{GstLAL}, \pycbc, \ac{MBTA} pipelines. The \ac{SPIIR} pipeline is performing similarly for {\hasremnant} but its performance is slightly lower for {\hasns}. {\hasmassgap}, on the other hand, has a \ac{TPR} of above $\sim 80 \%$ at $\sim 20 \%$ \ac{FPR}.

This paper presents the low-latency data products for O4 and their expected performance. Additional data products that expand on the current EM-Bright products for determining the likelihood of a kilonova are being developed that include predictions of mass ejecta for \ac{BNS} and \ac{NSBH} events, as well as peak magnitudes for corresponding kilonovae. We hope to make these data products public in the future.

\begin{acknowledgments}
We thank Varun Bhalerao for review of this paper. We thank the MBTA team for their contributions and for allowing the use of their pipeline data.

AT and MWC acknowledge support from the National Science Foundation with grant numbers PHY-2010970 and OAC-2117997. SSC and MC acknowledge support from the National Science Foundation with grant number PHY-2011334 and PHY-2308693. MC would also acknowledge support from NSF PHY-2219212. DC would like to acknowledge support from the NSF grants OAC-2117997 and PHY-1764464. EK, EM, GM would like to acknowledge support from NSF grant PHY-1764464. SM acknowledges support from JSPS Grant-in-Aid for Transformative Research Areas (A) No.~23H04891 and No.~23H04893. SA thanks the MITI CNRS for  their financial support. SG acknowledges NSF PHY-2110576. NA, PB, AB, PB, YKC, CM, and JC acknowledges NSF PHY-2207728. 
TD and VVO have received financial support from Xunta de Galicia (CIGUS Network of research centers), by European Union ERDF 
and by the ``María de Maeztu'' Units of Excellence program CEX2020-001035-M and the Spanish Research State Agency, and are supported by the research grant PID2020-118635GB-I00 from the Spanish Ministerio de Ciencia e Innovaci{\'o}n.

\end{acknowledgments}

\appendix

\begin{figure}[h!]
\centering
\includegraphics[width=0.6\linewidth]{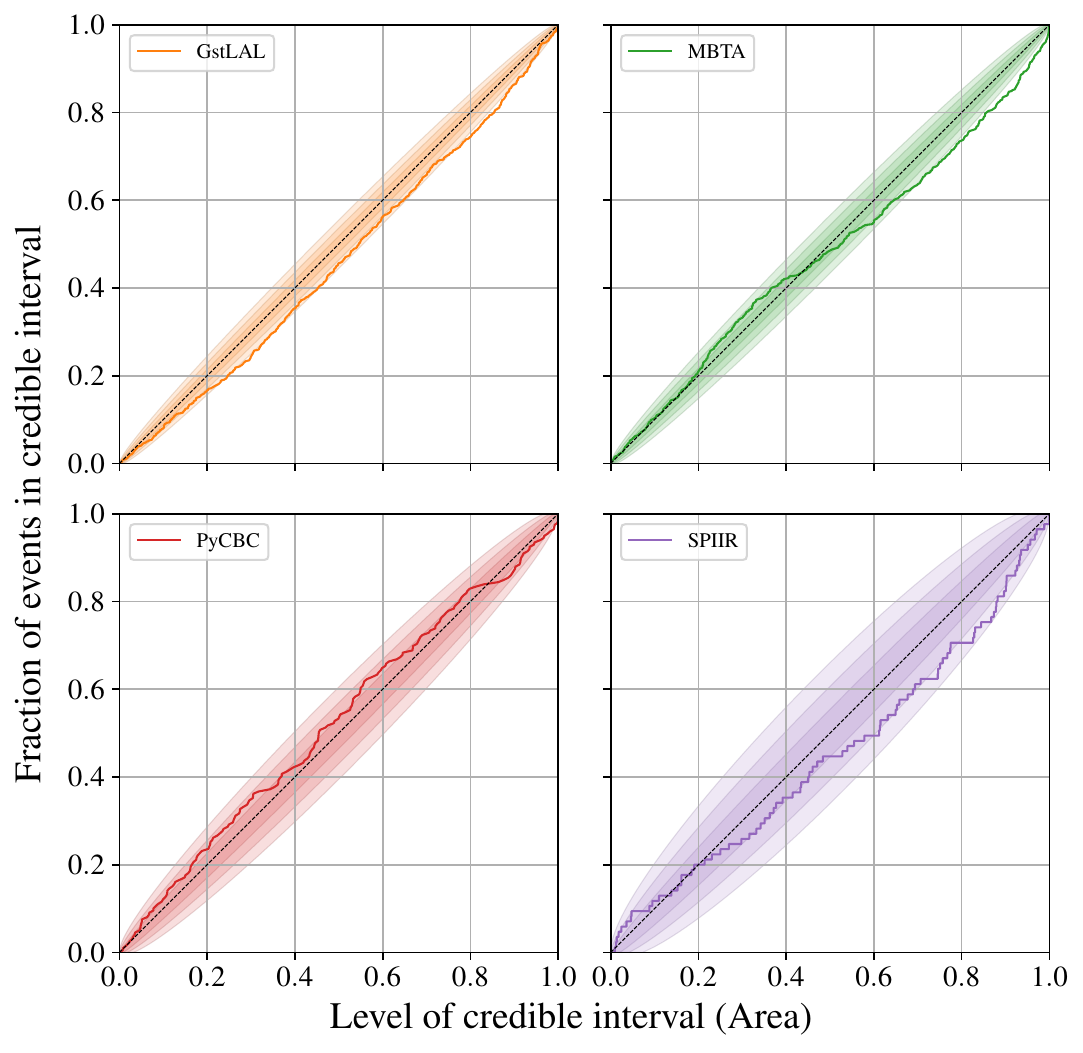}
\caption{\textit{P-P} plots showing \texttt{BAYESTAR} performance for pipelines that detected the preferred event. All the sky maps generated show that the performance of the sky maps are within the confidence bands.}
\label{fig:pp_bayestar_pipeline_grid}
\end{figure} 

\section{Sky map Performance}
\begin{figure}[h!]
\centering
\includegraphics[width=0.5\linewidth]{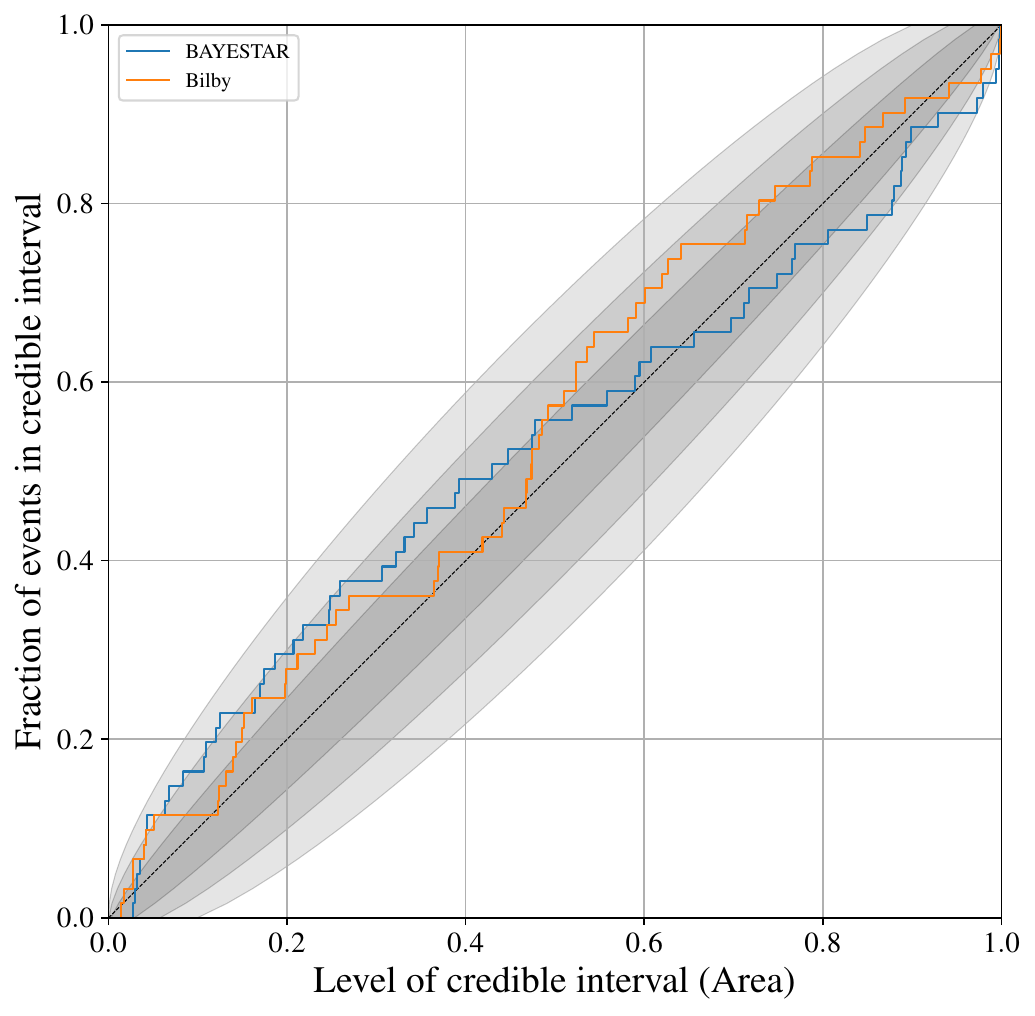}
\caption{\textit{P-P} plot comparing the performance of \texttt{Bilby} and \texttt{BAYESTAR} for \ac{BNS} injections likely to be the subject of follow-up, including the cuts on mass, spin, FAR, and number of interferometers as covered in Section \ref{subsec:skymapresults}. 
We see both \texttt{Bilby}'s and \texttt{BAYESTAR}'s performance is improved and within the credible intervals when including these cuts.}
\label{fig:pp_bilby_subset}
\end{figure}

The additional plots shown in this section are provided to demonstrate \texttt{Bilby} and \texttt{BAYESTAR} produce accurate sky localizations from injections found by each individual \ac{CBC} pipeline. Figure \ref{fig:pp_bayestar_pipeline_grid} shows the \texttt{BAYESTAR} performance of preferred events from each pipeline falls within the credible intervals besides minor deviations, and uses the same set of events and cuts as found in Figure \ref{fig:skymaps} and discussed in Section \ref{subsec:skymapresults}. To demonstrate performance for the events most likely to be subject to extensive follow-up, Figure \ref{fig:pp_bilby_subset} specifically presents only \ac{BNS} injections that pass the cuts applied in Figure \ref{fig:skymaps}. With these cuts we find the combined performance for both \texttt{Bilby} and \texttt{BAYESTAR} falls within for within credible intervals.

\clearpage


\bibliography{references}{}
\bibliographystyle{aasjournal}



\end{document}